\documentclass[preprint,showpacs,preprintnumbers,amsmath,amssymb]{revtex4}

\usepackage{graphicx}
\usepackage{epsfig}
\usepackage{amsmath}
\usepackage{graphics}

\newcommand{\fcoo}{{\bf q}, \omega} 
\newcommand{\coo}{{\bf r}, t} 
\newcommand{\intmom}{\int \frac{d{\bf p}}{(2\pi\hbar)^{3}}} 
 
\def\Vec#1{\mbox{\boldmath $#1$}}
\def\V#1{\mbox{\boldmath $#1$}}
\newcommand{\intxi}{\int d\Vec{\xi}}

\begin{document}

\title{Zero and First Sound in Normal Fermi Systems} 

\author{Shohei Watabe,$^{1}$ Aiko Osawa,$^{2}$ and Tetsuro Nikuni$^{3}$} 
\affiliation{
$^{1}$
Institute of Physics, Department of Physics, The University of Tokyo, Komaba 3-8-1 Meguro-ku, Tokyo, Japan, 153-8902
}
\affiliation{
$^{2}$
Department of Physics, Faculty of Science, 
Tokyo Institute of Technology, 
2-12-1 O-okayama, Meguro-ku, Tokyo, 
Japan, 152-8551}
\affiliation{
$^{3}$
Department of Physics, Tokyo University of Science, 
1-3 Kagurazaka, Shinjuku-ku, Tokyo, 
Japan, 162-8601}

%\date{}

\begin{abstract}
On the basis of a moment method, 
general solutions of a linearized Boltzmann equation for a normal Fermi system are investigated. 
In particular, we study the sound velocities and damping rates 
as functions of the temperature and the coupling constant. 
In the extreme limits of collisionless and hydrodynamic regimes, 
eigenfrequency of sound mode obtained from the moment equations 
reproduces the well-known results of zero sound and first sound. 
In addition, the moment method can describe crossover between those extreme limits at finite temperatures. 
Solutions of the moment equations also involve a thermal diffusion mode. 
From solutions of these equations, 
we discuss excitation spectra corresponding to the particle-hole continuum as well as collective excitations. 
We also discuss a collective mode in a weak coupling case. 
\end{abstract}


\maketitle

\section{Introduction}\label{introduction}

In discussing collective excitations of quantum degenerate gases, 
there arises a distinction between hydrodynamic and collisionless regimes. 
Let $\omega$ be a frequency of a collective mode and $\tau$ be a mean-collision time. 
The hydrodynamic regime is characterized by $\omega \tau \ll 1$, 
while the collisionless regime is characterized by $\omega \tau \gg 1$. 
The present paper is devoted to detail analyses of collective excitations 
in both limiting regimes and in the crossover regime. 
We deal with a sound propagation in a population balanced normal Fermi gas. 

%
%Let us give an overview of studies on collective modes in Fermi systems. 
%First, we review earlier theoretical studies with the Fermi liquid theory, and earlier experiments of liquid $^{3}$He 
%known to be an existent Fermi liquid at low temperatures.  
%After that, we give a review of recent studies in ultracold atomic gases. 

In general, 
the viscosity calculated in the hydrodynamic regime diverges at the absolute zero temperature ($T = 0$), 
and hence it was considered that sound could not propagate at $T=0$. 
In 1957, using the Fermi liquid theory~\cite{Landau19561957}, 
Landau predicted that 
a sound propagation could occur in liquid $^{3}$He even at very low temperatures 
owing to the mean-field interaction~\cite{Landau1957}. 
This new type of sound was called zero sound, 
which differs from first sound propagating because of a small dissipation achieved by local equilibrium. 

Khalatnikov and Abrikosov 
conducted detail calculations of the dispersion relation and derived the sound attenuation coefficient 
of the zero and first sound modes~\cite{Khalatnikov1958}. 
Abel {\it et. al.} confirmed the existence of zero sound in the liquid $^{3}$He~\cite{Abel1966}. 
They also observed a crossover between the zero and first sound modes, 
and measured temperature dependence of the sound velocity and of the sound attenuation coefficient. 
The theoretical prediction based on the Landau's Fermi liquid theory~\cite{Khalatnikov1958} 
agreed with the experimental data. 
Until now, investigations of liquid $^{3}$He have been conducted in detail, 
and the results are summarized in many text books~\cite{Helium3BOOK}. 
The Landau's Fermi liquid theory is also summarized in standard textbooks~\cite{FermiLiquidBOOK}. 

With the realization of the Bose-Einstein condensate 
as a turning point~\cite{BEC1995}, vigorous studies of ultracold atomic gases have been conducted. 
Ultracold atomic gases have flexibilities such as controllability of an interaction parameter using the Feshbach resonance.  
These systems open new windows to investigate phenomena that were 
difficult and impossible to study in the liquid helium and superconductors.

As mentioned below, 
a number of studies on the collective modes have been also reported in ultracold Fermi gases. 
Dipole oscillations were studied in collisionless and collisional regimes 
in two component $^{40}$K gases~\cite{Gensemer2001,DeMarco2002}. 
Experiments of the collective excitation in the BCS-BEC crossover regime have been performed 
using $^{6}$Li gases~\cite{Kinast2004,Bartenstein2004,Wright2007,Joseph2007}. 
In particular, the reference~\cite{Joseph2007} reported the sound velocity in the BCS-BEC crossover regime.

Collective excitations in Fermi gases 
have been also investigated theoretically. 
The dipole mode~\cite{Vichi1999} and the quadrupole mode~\cite{Vichi2000} were analyzed 
making use of the moment method. 
Zero sound with arbitrary spin was investigated~\cite{Yip1999}. 
Bruun {\it et. al.} studied collective modes in trapped gases 
extensively and intensively~\cite{Bruun1999,Bruun2001,Bruun2005A}. 
Tosi's group studied collective modes by solving the Boltzmann equation numerically~\cite{Toschi2002,Toschi2003,Akdeniz2003,Capuzzi2005}. 
They studied the dipole mode of two component trapped gases 
as a function of a collision rate in some situations~\cite{Toschi2002,Toschi2003}, 
and the crossover between zero and first sound modes in the cigar-shaped trap~\cite{Akdeniz2003,Capuzzi2005}. 
Recently, collective excitations in the unitarity limit were studied~\cite{Massignan2005,Bruun2007,Taylor20072008}.  

As noted earlier, after a publication of the path-breaking work by Landau~\cite{Landau1957}, 
a classic paper by Khalatnikov and Abrikosov studied the crossover 
between zero sound and first sound~\cite{Khalatnikov1958}. 
They treated a mean-collision time as temperature-dependent, 
but approximated other quantities by those at $T = 0$. 
When the system is in the collisional regime at finite temperatures, 
a sound velocity within this analysis reaches that of first sound evaluated at $T = 0$. 
This simple approximation is appropriate as long as we restrict discussion to the Landau's Fermi liquid theory, 
since this theory focuses on quasiparticles at sufficiently low temperatures. 
In the experiments in liquid $^{3}$He~\cite{Abel1966}, however, 
the temperature dependence of first sound velocity has been observed, although it was small. 
In usual atomic gases, furthermore, 
first sound has a significant temperature dependence, 
and hence the simple treatment mentioned above is invalid at high temperatures. 
It is therefore necessary to discuss the crossover between the zero and first sound modes 
with a more efficient treatment valid at wide range of temperatures.

Brooker and Sykes attempted to analyze the general solution of the linearized Boltzmann equation to investigate 
the crossover of sound propagation~\cite{Brooker1970}. 
They expanded a deviation from local equilibrium in terms of spherical harmonic functions, 
and introduced different relaxation times for different spherical harmonics. 
They, however, used an approximation only appropriate in the low temperature regime, 
and hence the analysis was not appropriate at high temperatures. 
They, furthermore, could not obtain the explicit solution because of the computational difficulty at that time, 
although they gave an equation to be solved. 

Although the physics on the crossover between the zero and first sound modes 
has been understood to some extent, some issues shown above still remain. 
With the developments of recent experimental techniques in ultracold atomic gases and of theoretical methods, 
it is meaningful to revisit the study of sound mode in Fermi systems with a modern approach. 
In the present paper, 
we investigate the crossover between zero sound and first sound 
over wide parameter ranges with a single theoretical framework. 
For this purpose, we analyze the general solution of the linearized Boltzmann equation using the moment method. 

The moment method is suitable for describing the crossover of collective excitations 
between collisionless and hydrodynamic regimes. 
So far in cold atomic gases, the moment method has been used to study characteristic collective excitations in trapped systems, 
such as monopole, dipole, quadrupole and scissors modes~\cite{Odelin1999A,Khawaja2000,Odelin1999B,Nikuni2002A}. 
With a use of same technique, 
collective modes in atomic gases with internal degrees of freedom were also studied~\cite{Nikuni2002B,Endo2008}.  
The moment equation for the uniform system, however, 
has not been solved to study sound modes in quantum many-body systems. 
This is one of the new points in the present paper.

In the Landau's Fermi liquid theory, central players are quasiparticle, and hence 
theoretical studies have been done in the very low temperature regime. 
Those studies are based on the Landau Boltzmann equation for quasiparticles. 
In dilute quantum gases, on the other hand, real atoms are central players. 
These systems are described by the Boltzmann equation for real particles. 
Although the Boltzmann equation has the same form as the Landau Boltzmann equation, 
this equation is applicable up to the high-temperature Maxwell-Boltzmann gas regime. 
Studies of sound mode in ultracold atomic gases thus need a single theory which can be applied up to high temperatures. 
The present study can solve those issues.

The contents of the present paper are summarized as follows: 
(a) The spectrum of the sound mode obtained from the general solution 
shows the crossover between zero sound and first sound.  
In collisionless and collisional regimes, 
the sound velocities reproduce the results calculated in each limiting regime. 
This method also offers the frequency of first sound with the temperature dependence. 
This result cannot be obtained by a standard approach such as given in Ref.~\cite{Khalatnikov1958}. 
(b) The results of the moment method include a thermal diffusion mode. 
(c) The moment method reproduces not only a collective mode of the sound propagation, 
but also the particle-hole continuum. 
(d) In a weak coupling case, the crossover between zero sound and first sound has a different character from 
the crossover in a strong coupling case.

The present paper is organized as follows. 
Section~\ref{moment} deals with one of the main topics of the present paper. 
Making use of the moment method, we will derive moment equations. 
In Sec.~\ref{first}, we calculate the sound velocity and damping rate of first sound in the hydrodynamic regime. 
Section~\ref{zero} gives detailed analysis of zero sound in the low temperature regime. 
In Sec.~\ref{results}, we will show results of moment equations, 
and analyze the crossover between zero sound and first sound. 
Section~\ref{discussion} is devoted to discussion. 
Section~\ref{summary} gives summary and conclusion. 
We devote Appendices~\ref{chapman} and~\ref{relaxation} to derive relaxation times. 
In Appendix~\ref{chapman}, we calculate transport coefficients 
in the hydrodynamic regime based on the Chapman-Enskog method. 
In Appendix~\ref{relaxation}, we evaluate relaxation times associated with the transport coefficients. 
We compare the mean collision rate with these relaxation rates: the viscous relaxation rate and the thermal conductivity relaxation rate. 
Appendix~\ref{RPA} describes a standard analysis of the random-phase approximation.

\section{Linearized Boltzmann Equation and Moment Equation}\label{moment}

In this paper, we consider two component atomic Fermi gas interacting with $s$-wave scattering. 
We assume a population balanced gas of two spin components with the same mass $m$. 
The equation of motion for distribution functions within a semiclassical approximation is described by the following Boltzmann equation: 
\begin{align}
\frac{\partial f_{\sigma}({\bf p}, \coo)}{\partial t}
+ 
\frac{\bf p}{m}\cdot \nabla_{\bf r} f_{\sigma}({\bf p}, \coo)
-
\nabla_{\bf r}U_{\sigma}(\coo)
\cdot
\nabla_{\bf p}f_{\sigma}({\bf p}, \coo)
= 
{\mathcal I}_{\rm coll}
[f_{\sigma}], 
\label{Boltzmann}
\end{align}
where an index $\sigma =\{ \uparrow, \downarrow\}$ represents spin component. 
$U_{\sigma}(\coo)$ is the contribution of a mean-field interaction 
given by $U_{\sigma}(\coo) = g n_{-\sigma}(\coo)$. 
$n_{\sigma}(\coo)$ represents the local density. 
The interaction strength $g$ is given by $g = 4\pi\hbar^{2}a/m$, 
and $a$ is the $s$-wave scattering length. 
The collision integral ${\mathcal I}_{\rm coll} [f_{\sigma}]$ on the right hand side of Eq. (\ref{Boltzmann}) is given by 
\begin{align}
{\mathcal I}_{\rm coll} [f_{\sigma}(1)] 
=& 
\frac{2\pi g^{2}}{\hbar}
\int \frac{d{\bf p}_{2}}{(2\pi\hbar)^{3}}
\int \frac{d{\bf p}_{3}}{(2\pi\hbar)^{3}}
\int d{\bf p}_{4}
\delta ({\bf p}_{1} + {\bf p}_{2} - {\bf p}_{3} - {\bf p}_{4})
\delta 
\left (
\frac{p_{1}^{2}}{2m} + \frac{p_{2}^{2}}{2m} 
- \frac{p_{3}^{2}}{2m} - \frac{p_{4}^{2}}{2m}
\right ) 
\nonumber
\\
&  
\times 
\left [ 1- f_{\sigma}(1) \right]
\left [ 1- f_{- \sigma}(2) \right] 
f_{- \sigma}(3)
f_{\sigma}(4)
- 
f_{\sigma}(1)
f_{-\sigma}(2)
\left [ 1- f_{-\sigma}(3) \right] 
\left [ 1- f_{\sigma}(4) \right],   
\label{Collision}
\end{align}
where $f_{\sigma}(i)$ is $f_{\sigma}(i) \equiv f_{\sigma}({\bf p}_{i}, {\bf r}, t)$.

We shall linearize the distribution function 
around static equilibrium (denoted by $f_{\sigma}^{0} ({\bf p}, {\bf r})$), 
using $f_{\sigma}({\bf p}, {\bf r}, t) 
= 
{f}_{\sigma}^{0} ({\bf p}, {\bf r}) + 
\delta f_{\sigma} ({\bf p}, {\bf r}, t) 
= 
{f}_{\sigma}^{0} ({\bf p}, {\bf r}) + 
\delta \tilde{f}_{\sigma} ({\bf p}, {\bf r}, t)
+ 
\delta f'_{\sigma} ({\bf p}, {\bf r}, t)$. 
Here, $\delta \tilde{f}_{\sigma} ({\bf p}, {\bf r}, t)$ is the local equilibrium distribution 
function linearized around static equilibrium 
$
\tilde{f}_{\sigma} ({\bf p}, {\bf r}, t) 
= 
f_{\sigma}^{0} ({\bf p}, {\bf r}) 
+
\delta \tilde{f}_{\sigma} ({\bf p}, {\bf r}, t)$, 
and $\delta f'({\bf p}, {\bf r}, t)$ denotes departure from local equilibrium. 

The local equilibrium distribution is determined 
by the condition ${\mathcal I}_{\rm coll} = 0$, and is given by 
\begin{align}
\tilde{f}_{\sigma} ({\bf p}, \coo) 
= 
\frac{1}{
\exp{
\left \{
\tilde{\beta}(\coo)\left [{\bf p} -  m \overline{\bf v} (\coo) \right ]^{2}/2m
\right \}
}
z_{\sigma}^{-1}(\coo) 
+ 1
},  
\label{local-distribution}
\end{align}
where the local fugacity $z_{\sigma}(\coo)$ is 
\begin{align}
z_{\sigma}(\coo) = 
\exp{ 
\left \{
\tilde{\beta}(\coo) \left[\tilde{\mu}_{\sigma}(\coo) -g\tilde{n}_{-\sigma}(\coo) \right]
\right \}
}.  
\label{local-fugacity}
\end{align}
$\tilde{\beta}(\coo)$ is a local temperature $\tilde{\theta}(\coo) = k_{\rm B}\tilde{T}(\coo) = 1/\tilde{\beta}({\bf r}, t)$, 
$\tilde{n}_{\sigma}(\coo)$ is a local density, 
$\mu_{\sigma}(\coo)$ is a local chemical potential, and $\overline{\bf v}(\coo)$ is local velocity. 
These local variables depend on position and time. 

It is convenient to write fluctuations of the distribution function around 
static equilibrium as 
\begin{align} 
\delta f_{\sigma}({\bf p}, {\bf r}, t) 
= 
\frac{\partial f_{\sigma}^{0}}{\partial \varepsilon_{\sigma}^{0}}
\nu_{\sigma}({\bf p}, {\bf r}, t). 
\label{asterisk} 
\end{align} 
The factor $\nu_{\sigma}({\bf p}, {\bf r}, t)$ is the averaged extra energy of particles around equilibrium~\cite{ZimanBOOK}. 
Writing $\delta \tilde{f}_{\sigma}({\bf p}, {\bf r}, t)$ and 
$\delta f_{\sigma}'({\bf p}, {\bf r}, t)$ as 
\begin{align}
\delta \tilde{f}_{\sigma}({\bf p}, {\bf r}, t) 
= 
\frac{\partial f_{\sigma}^{0}}{\partial \varepsilon_{\sigma}^{0}} 
\tilde{\nu}_{\sigma}({\bf p}, {\bf r}, t), 
\qquad 
\delta f'_{\sigma}({\bf p}, {\bf r}, t) 
= 
\frac{\partial f_{\sigma}^{0}}{\partial \varepsilon_{\sigma}^{0}} 
\delta \nu_{\sigma}({\bf p}, {\bf r}, t), 
\end{align}
one also has 
$
\nu_{\sigma}({\bf p}, {\bf r}, t) = 
\tilde{\nu}_{\sigma}({\bf p}, {\bf r}, t)+\delta \nu_{\sigma}({\bf p}, {\bf r}, t)
$. 

Using Eq. (\ref{asterisk}) in the Boltzmann equation (\ref{Boltzmann}), 
one obtains the following equation: 
\begin{align} 
\frac{\partial f_{\sigma}^{0}}{\partial \varepsilon_{\sigma}^{0}}
\left [ 
\frac{\partial \nu_{\sigma} ({\bf p}, {\bf r}, t)}{\partial t} 
+ 
\frac{{\bf p}}{m}\cdot \nabla \nu_{\sigma} ({\bf p}, {\bf r}, t) 
- \nabla g\delta n_{-\sigma}({\bf r}, t) \cdot \frac{{\bf p}}{m}
\right ]
= 
{\mathcal I}_{\rm coll} [f_{\sigma}]. 
\end{align}
We shall apply a relaxation time approximation to the collision integral. 
The relaxation time $\tau$ is a characteristic time with which a system reach local equilibrium. 
In this approximation, the collision integral can be reduced to 
\begin{eqnarray}
{\mathcal I}_{\rm coll}[f_{\sigma}] 
&=&
-\frac{f_{\sigma}-\tilde{f}_{\sigma}}{\tau}
= 
-\frac{1}{\tau}
\frac{\partial f_{\sigma}^{0}}{\partial \varepsilon_{\sigma}^{0}}
\delta \nu_{\sigma}. 
\end{eqnarray}

We now linearize the local equilibrium quantities as 
$\tilde{\theta} ({\bf r},t) = \theta^{0} + \delta \theta ({\bf r}, t)$, 
$\tilde{\mu}_{\sigma} ({\bf r}, t) = \mu_{\sigma}^{0} + \delta \mu_{\sigma} ({\bf r}, t)$, 
and 
$\overline{\bf v} ({\bf r}, t) = {\bf v}^{0}  + \delta {\bf v} ({\bf r}, t) = \delta {\bf v} ({\bf r}, t)$, 
where $\theta^{0}= k_{\rm B}T^{0}$, 
chemical potential $\mu^{0}$, and velocity ${\bf v}^{0} = {\bf 0}$ represent 
static equilibrium. 
The linearized local equilibrium distribution function is then given by 
\begin{eqnarray}
\tilde{\nu}_{\sigma}({\bf p}, {\bf r}, t) 
&=&
a_{\sigma}({\bf r}, t) +
{\bf b}({\bf r}, t) \cdot{\bf p} + c({\bf r}, t)  p^{2}, 
\end{eqnarray}
where 
\begin{eqnarray}
a_{\sigma}({\bf r}, t)
&\equiv& 
-\left [
\beta_{0} g n_{-\sigma}^{0}
-\beta_{0}\mu_{\sigma}^{0}
\right ]
\delta \theta({\bf r}, t)
- 
\delta \mu_{\sigma}({\bf r}, t)
+ g\delta n_{-\sigma}({\bf r}, t), 
\\
{\bf b}({\bf r}, t) &\equiv& 
- \delta{\bf v} ({\bf r}, t), 
\label{B-v-relation}
\\
c({\bf r}, t)  &\equiv& 
-\frac{\beta_{0}}{2m}
\delta\theta ({\bf r}, t).  
\end{eqnarray}

We now look for the plane wave solution of the linearized Boltzmann equation, representing as 
\begin{align}
\nu_{\sigma} ({\bf p}, {\bf r}, t) = 
\nu_{\sigma}({\bf p}, {\bf q}, \omega) e^{i({\bf q}\cdot{\bf r} -\omega t)}, 
\qquad 
\delta n_{\sigma} ({\bf r}, t) = 
\delta n_{\sigma}({\bf q}, \omega) e^{i({\bf q}\cdot{\bf r} -\omega t)}, 
\end{align} 
where resultant functions $\tilde{\nu}_{\sigma}$ and $\delta\nu_{\sigma}$ 
are also written as 
\begin{align}
&
\nu_{\sigma}({\bf p}, {\bf q}, \omega) = 
\tilde{\nu}_{\sigma}({\bf p}, {\bf q}, \omega)
+\delta\nu_{\sigma}({\bf p}, {\bf q}, \omega), 
\label{Equation83}
\\
&
\tilde{\nu}_{\sigma}({\bf p}, {\bf q}, \omega) 
= 
a_{\sigma}({\bf q}, \omega) + {\bf b}({\bf q}, \omega) \cdot {\bf p} 
+ c({\bf q}, \omega) p^{2}. 
\label{Equation84}
\end{align}
The linearized Boltzmann equation with the relaxation time approximation is thus given by 
\begin{eqnarray}
\frac{\partial f_{\sigma}^{0}}{\partial \varepsilon_{\sigma}^{0}}
\left [ 
\left (
\omega -\frac{{\bf p}\cdot{\bf q}}{m}
\right )
\nu_{\sigma}({\bf p}, {\bf q}, \omega)
+ 
\frac{{\bf p}\cdot{\bf q}}{m}
g \delta n_{-\sigma} ({\bf q}, \omega)
\right ]
= 
-i 
\frac{1}{\tau}
\frac{\partial f_{\sigma}^{0}}{\partial \varepsilon_{\sigma}^{0}}
\delta \nu_{\sigma}({\bf q}, \omega). 
\label{Equation82}
\end{eqnarray} 
Here, the density fluctuation $\delta n_{\sigma}({\bf q}, \omega)$ is 
\begin{eqnarray}
\delta n_{\sigma} ({\bf q}, \omega)
&=&
a_{\sigma}({\bf q}, \omega) W_{\sigma, 0}
+ 
c({\bf q}, \omega) W_{\sigma, 2}
+ 
\int \frac{d{\bf p}}{(2\pi\hbar)^{3}}
\frac{\partial f_{\sigma}^{0}}{\partial \varepsilon_{\sigma}^{0}}
\delta \nu_{\sigma} ({\bf q}, \omega), 
\label{delta-density}
\end{eqnarray}
where we define 
\begin{align}
W_{\sigma, n} \equiv \intmom \frac{\partial f_{\sigma}^{0}}{\partial \varepsilon_{\sigma}^{0}}p^{n}. 
\end{align}
From Eqs. (\ref{Equation83})-(\ref{delta-density}),
the linearized Boltzmann equation is reduced to 
\begin{align}
\frac{\partial f_{\sigma}^{0}}{\partial \varepsilon_{\sigma}^{0}}
\left \{ 
\left (
\omega -\frac{{\bf p}\cdot{\bf q}}{m}
\right )
\left [
a_{\sigma}({\bf q}, \omega)
+ {\bf b}({\bf q}, \omega)\cdot{\bf p}
+ c({\bf q}, \omega)p^{2}
+ \delta \nu_{\sigma}({\bf q}, \omega)
\right ]
\right . 
&
\nonumber
\\
\left .
+ 
\frac{{\bf p}\cdot{\bf q}}{m}
g 
\left [
a_{-\sigma}({\bf q}, \omega)
W_{-\sigma, 0}
+ 
c({\bf q}, \omega) W_{-\sigma, 2}
+ 
\int \frac{d{\bf p}}{(2\pi\hbar)^{3}}
\frac{\partial f_{-\sigma}^{0}}{\partial \varepsilon_{-\sigma}^{0}}
\delta \nu_{-\sigma}({\bf q}, \omega)
\right ]
\right \}
&
= 
\displaystyle{
-i\frac{1}{\tau}
\frac{\partial f_{\sigma}^{0}}{\partial \varepsilon_{\sigma}^{0}}
\delta \nu_{\sigma}
}. 
\label{linearized+deltamu}
\end{align}

We now discuss the general solution of the linearized Boltzmann equation, 
a main topic of the present paper. 
As derived in the above, the linearized Boltzmann equation 
with the averaged extra energy around the static equilibrium $\nu_{\sigma}({\bf p})$ is reduced to 
\begin{equation}
\frac{\partial f_{\sigma}^{0}}{\partial \varepsilon_{\sigma}^{0}}
\left [ 
\left (
\omega -\frac{{\bf p}\cdot{\bf q}}{m}
\right )
\nu_{\sigma}({\bf p}) 
+ 
\frac{{\bf p}\cdot{\bf q}}{m}
g \delta n_{-\sigma} 
\right ]
= 
-\frac{1}{\tau}
\frac{\partial f_{\sigma}^{0}}{\partial \varepsilon_{\sigma}^{0}}
\left [\nu_{\sigma} - \left ( a_{\sigma} + {\bf b}\cdot {\bf p} + cp^{2} \right )\right ]. 
\label{Equation112}
\end{equation} 
We do not write explicitly ${\bf q}$, and $\omega$ in $\nu_{\sigma}$ and $\delta n_{\sigma}$, 
since these dependences are not important for further calculation. 
We shall use the viscous relaxation time given in Eq. (\ref{tau-eta}) (derived in Appendix~\ref{chapman} and ~\ref{relaxation}) 
as the relaxation time $\tau$, 
because the density fluctuation is the most strongly coupled to the 
viscous relaxation. 
This approximation to the collision integral is a good one in the vicinity of local equilibrium. 
In the collisionless regime $\omega\tau \gg 1$, 
this approximation can describe zero sound, 
because the collision term can be neglected owing to the large value of $\tau$. 
We remark that the relaxation time evaluated by a small correction from 
static and local equilibrium could be quantitatively different from 
the actual relaxation time in the collisionless regime 
$\omega\tau \gg 1$ and also in the intermediate regime $\omega\tau \approx 1$.

We expand the fluctuation in terms of the spherical harmonics as 
\begin{eqnarray}
\nu_{\sigma}({\bf p}) 
\equiv 
\sum\limits_{l = 0}^{\infty}\sum\limits_{m = -l}^{l}
\nu_{\sigma, l}^{m} (p) P_{l}^{m}(\cos{\theta})e^{im\phi}. 
\end{eqnarray} 
Multiplying Eq. (\ref{Equation112}) by $e^{-im'\phi}$ 
and integrating it over $\phi$, 
we have the following linearized Boltzmann equation: 
\begin{align}
&
\sum\limits_{l = 0}^{\infty}
\frac{\partial f_{\sigma}^{0}}{\partial \varepsilon_{\sigma}^{0}}
\left [
\left ( 
\omega -\frac{pq}{m}\cos{\theta} 
\right )
\nu_{\sigma,l}^{m} P_{l}^{m} (\cos{\theta})
\right ]
+ 
\frac{\partial f_{\sigma}^{0}}{\partial \varepsilon_{\sigma}^{0}}
\left (
\frac{pq}{m} \cos{\theta} 
\right )
g \delta n_{-\sigma}\delta_{m, 0}
\nonumber
\\
= &
-\frac{i}{\tau}
\frac{\partial f_{\sigma}^{0}}{\partial \varepsilon_{\sigma}^{0}}
\left [\sum\limits_{l = 0}^{\infty}\nu_{\sigma}^{m}P_{l}^{m}(\cos{\theta}) - \left ( a_{\sigma} + {\bf b}\cdot {\bf p} + cp^{2} \right ) \delta_{m,0}\right ]. 
\label{Equation116}
\end{align} 

One finds that only the mode with $m=0$ is coupled to the mean-field potential. 
This is due to the isotropic interaction. 
This mode $(m=0)$ corresponds to the longitudinal wave. 
In anisotropic interactions, 
there also exists the mode with $m \neq 0$, 
such as transverse zero sound with $m = 1$. 
Since we consider the crossover from the longitudinal zero sound to the longitudinal first sound, 
we only take the mode with $m = 0$. 
Let us use the notations 
$\nu_{\sigma, l}^{m = 0}(p) \equiv \nu_{\sigma, n}(p)$, and 
$P_{l}^{m=0} (\cos{\theta}) \equiv P_{l} (\cos{\theta})$, for simplicity. 
It is also useful to define the following moment: 
\begin{equation}
\label{ }
\langle p^{n} \nu_{\sigma, l}\rangle 
\equiv \int \frac{d{\bf p}}{(2\pi\hbar)^{3}}
\frac{\partial  f_{\sigma}^{0}}{\partial \varepsilon_{\sigma}^{0}}
p^{n} \nu_{\sigma, l}(p). 
\end{equation}
The density fluctuation is expressed as $\delta n_{\sigma} = \langle \nu_{\sigma, 0}\rangle$. 

Multiplying Eq. (\ref{Equation116}) by $p^{n} P_{l'}( \cos{\theta} )$ 
and integrating over $\theta$ and ${\bf p}$, 
we obtain the moment equation given by 
\begin{align} 
&
\omega
\langle p^{n}\nu_{\sigma, l}\rangle
-\frac{l}{2l-1} \frac{q}{m} \langle p^{n+1}\nu_{\sigma, l-1}\rangle
-\frac{l+1}{2l+3} \frac{q}{m} 
\langle p^{n+1}\nu_{\sigma, l+1}\rangle
+g\frac{q}{m}W_{\sigma,n+1}  
\langle \nu_{-\sigma, 0}\rangle \delta_{l,1} 
\nonumber 
\\
=
&
-\frac{i}{\tau} 
\langle p^{n}\nu_{\sigma, l}\rangle
+\frac{i}{\tau}
\left ( 
a_{\sigma}W_{\sigma, n}
+ c W_{\sigma, n+2}
\right )\delta_{l, 0}
+\frac{i}{\tau}
b 
W_{\sigma, n+1}
\delta_{l, 1}. 
\label{CoefficientMatrix}
\end{align}
We have made use of an orthogonality relation 
\begin{eqnarray}
\int_{0}^{\pi}d\theta \sin{\theta}
P_{l}(\cos{\theta})P_{l'}(\cos{\theta}) 
&=& 
\frac{2}{2l + 1}\delta_{l, l'}, 
\end{eqnarray}
and a recurrence formula for the Legendre polynomials 
\begin{eqnarray}
\cos{\theta} P_{l}(\cos{\theta}) &= &
\frac{l+1}{2l + 1} P_{l+1}(\cos{\theta}) + \frac{l}{2l+1}P_{l-1}(\cos{\theta}). 
\end{eqnarray}
 
The moments associated with $p^{0}P_{0}(\cos{\theta})$, $pP_{1}(\cos{\theta})$ and $p^{2}P_{0}(\cos{\theta})$ 
correspond to number of particles, momentum, and the energy, respectively. 
The collision integral vanishes when we take these moments, because of the conservation law. 
Equations determining coefficients $a_{\sigma}$, $b$, and $c$ are then given by 
\begin{eqnarray}
\langle \nu_{\sigma, 0} \rangle  - a_{\sigma}W_{\sigma, 0} - c W_{\sigma, 2} &= 0, 
\label{CNMoment}
\\
\sum\limits_{\sigma} \left ( \langle p\nu_{\sigma, 1} \rangle - b W_{\sigma, 2} \right ) & = 0,  
\label{CMMoment}
\\
\sum\limits_{\sigma}
\left ( \langle p^{2} \nu_{\sigma, 0} \rangle - a_{\sigma} W_{\sigma, 2} - c W_{\sigma, 4} \right )
& = 0. 
\label{CEMoment}
\end{eqnarray}
We used an assumption that directions of the velocity and of the sound propagation are 
parallel ${\bf b} \parallel {\bf k}$, where ${\bf b}$ is related to the velocity through Eq. (\ref{B-v-relation}). 
As a result, one obtains coefficients given by 
\begin{align}
a_{\sigma}
= &   
\frac{1}{W_{\sigma, 0}}
\langle \nu_{\sigma, 0} \rangle
- 
\frac{1}{\Theta}
\frac{W_{\sigma, 2}}{W_{\sigma, 0}}
\sum\limits_{\sigma'}
\langle p^{2} \nu_{\sigma', 0} \rangle 
+ 
\frac{1}{\Theta}
\frac{W_{\sigma, 2}}{W_{\sigma, 0}}
\sum\limits_{\sigma'}
\frac{W_{\sigma', 2}}{W_{\sigma', 0}}
\langle \nu_{\sigma', 0} \rangle,  
\\
b = &
\frac{\langle p \nu_{\uparrow, 1} \rangle + \langle p \nu_{\downarrow, 1} \rangle }
{W_{\uparrow, 2} + W_{\downarrow, 2}}, 
\\
c = &
\frac{1}{\Theta}
\left ( 
\sum\limits_{\sigma}
\langle p^{2} \nu_{\sigma, 0} \rangle 
-
\sum\limits_{\sigma}
\frac{W_{\sigma, 2}}{W_{\sigma, 0}}
\langle \nu_{\sigma, 0} \rangle 
\right ), 
\end{align}
where 
$
\Theta \equiv 
\sum\limits_{\sigma}
\left (
W_{\sigma, 4}
-
W_{\sigma, 2}^{2} / W_{\sigma, 0}
\right )
$. 
Finally, we obtain the following moment equation: 
\begin{align}
&
\omega
\langle p^{n}\nu_{\sigma, l}\rangle
-\frac{l}{2l-1} \frac{q}{m} \langle p^{n+1}\nu_{\sigma, l-1}\rangle
-\frac{l+1}{2l+3} \frac{q}{m} 
\langle p^{n+1}\nu_{\sigma, l+1}\rangle
+g\frac{q}{m}W_{\sigma,n+1}  
\langle \nu_{-\sigma, 0}\rangle \delta_{l,1}
\nonumber 
\\
=& 
-
\frac{i}{\tau}
\langle p^{n} \nu_{\sigma,l} \rangle 
\nonumber 
\\
&
+ 
\frac{i}{\tau}
\left [
\frac{W_{\sigma,n}}{W_{\sigma,0}}
+ 
\frac{1}{\Theta}
\left ( 
\frac{ W_{\sigma,n} W_{\sigma,2}^{2} }{W_{\sigma,0}^{2} }
- 
\frac{ W_{\sigma,n+2} W_{\sigma,2} }{ W_{\sigma,0} }
\right ) 
\right ]
\langle
\nu_{\sigma, 0} 
\rangle
\delta_{l,0} 
\nonumber 
\\
&
+ 
\frac{i}{\tau}
\frac{1}{\Theta}
\left ( 
\frac{ W_{\sigma,n} W_{\sigma,2} W_{-\sigma,2}}
{W_{\sigma,0} W_{-\sigma,0} }
- 
\frac{ W_{\sigma,n+2} W_{-\sigma,2} }{ W_{-\sigma,0} }
\right ) 
\langle
\nu_{-\sigma, 0} 
\rangle
\delta_{l,0} 
\nonumber 
\\
&
+ 
\frac{i}{\tau}
\frac{1}{\Theta}
\left (
W_{\sigma, n+2}
- 
\frac{ W_{\sigma,2} W_{\sigma, n} }{W_{\sigma,0} }
\right ) 
\left ( 
\langle
p^{2} \nu_{\uparrow, 0} 
\rangle
+
\langle
p^{2} \nu_{\downarrow, 0} 
\rangle
\right )
\delta_{l,0} 
\nonumber 
\\
&
+ 
\frac{i}{\tau}
\frac{W_{\sigma,n+1}}{W_{\uparrow,2}+ W_{\downarrow, 2}}
\left ( 
\langle
p \nu_{\uparrow, 1} 
\rangle
+
\langle
p \nu_{\downarrow, 1} 
\rangle
\right )
\delta_{l,1}. 
\label{MomentEquation}
\end{align} 

One can obtain the eigenmode by solving this eigenvalue problem; 
however, equations are not closed even if higher moments are taken into account. 
We shall truncate an equation at sufficiently high moment, 
which does not affect the spectrum of the collective mode of interest. 
%We shall solve the secular equation of the linearized Boltzmann equation. 
Note that this equation is not the same one derived in Ref.~\cite{Brooker1970}.  
The equation (\ref{MomentEquation}) is much simpler than that in Ref.~\cite{Brooker1970}. 
We do not use many relaxation times as in Ref.~\cite{Brooker1970}, 
but a single relaxation time is introduced. 
Reference~\cite{Brooker1970} added an extra equation to make a closed set of equations, 
but we do not need an extra equation. 
In next two sections, 
we grasp sound velocities and damping rates in the two limiting regimes: hydrodynamic and collisionless regimes. 

\section{First Sound}\label{first}
We solve the linearized Boltzmann equation in Eq. (\ref{linearized+deltamu}) 
in the hydrodynamic regime in the present section. 
When we take the zeroth, first, and second moments 
of the Boltzmann equation, 
the collision integral vanishes because of conservation laws. 
As shown in Eqs. (\ref{velocityEq-with-viscosity}) and (\ref{energydensityEq-with-heatcurent}) in Appendix~\ref{chapman}, 
one can obtain a closed set of hydrodynamic equations including dissipative terms. 
The hydrodynamic equations in terms of the moments are written as 
\begin{align}
&
\, 
\omega
\left [ 
a_{\sigma}(\fcoo) W_{\sigma, 0}
+ 
c(\fcoo) W_{\sigma, 2}
\right ] 
= 
\frac{{\bf b}(\fcoo)\cdot{\bf q}}{3m}W_{\sigma,2}, 
\\
&
\sum\limits_{\sigma}
\left \{ 
\omega{\bf b}(\fcoo) W_{\sigma,2}
-
a_{\sigma}(\fcoo) \frac{W_{\sigma,2}}{m}{\bf q}
-
c(\fcoo)\frac{W_{\sigma,4}}{m}{\bf q}
\right. 
\nonumber
\\
&
\left .
\qquad \qquad  \qquad
+ 
g \frac{W_{\sigma,2}}{m}
\left [
a_{-\sigma}(\fcoo) W_{-\sigma,0} 
+ 
c(\fcoo)W_{-\sigma,2}
\right ]
{\bf q}
\right \}
-
i4\eta q^{2} 
{\bf b}({\bf q}, \omega) =0, 
\\
&
\omega 
\left [
a_{\uparrow}(\fcoo) W_{\uparrow,2}
+ 
a_{\downarrow}(\fcoo) W_{\downarrow,2}
+ 
c(\fcoo)
(W_{\uparrow,4}+W_{\downarrow,4})
\right ] 
\nonumber
\\
&
\qquad \qquad \qquad
-
\frac{{\bf b}({\bf q}, \omega)\cdot{\bf q}}{3m}
(W_{\uparrow,4}+W_{\downarrow,4})
- 
4i\kappa m^{2} c({\bf q}, \omega) T q^{2}
=0. 
\end{align} 
Relations $\delta \theta (\fcoo) = k_{\rm B} \delta T(\fcoo)$ and 
${\bf b} ({\bf q}, \omega) = - \overline{\bf v}({\bf q}, \omega)$ are used, 
and the velocity $\overline{\bf v} ({\bf r}, t)$ is assumed to be parallel to a vector ${\bf q}$. 

The above equations can be written in terms of physical quantities: 
the density $\delta n_{\sigma}({\bf q}, \omega)$, 
the velocity $\overline{\bf v}({\bf q}, \omega)$ and the energy $\delta E({\bf q}, \omega)$, 
whose quantities are given by 
\begin{align}
\delta n_{\sigma} ({\bf q}, \omega) = & a_{\sigma}({\bf q}, \omega) W_{\sigma,0} + c({\bf q}, \omega) W_{\sigma, 2},  
\\
\overline{\bf v}({\bf q}, \omega) = & - {\bf b} ({\bf q}, \omega),  
\\
\delta E ({\bf q}, \omega) = & a_{\uparrow}({\bf q}, \omega) W_{\uparrow, 2} 
+ a_{\downarrow}({\bf q}, \omega) W_{\downarrow, 2} +  c({\bf q}, \omega) ( W_{\uparrow, 2} + W_{\downarrow, 2} ). 
\end{align} 
Density fluctuations of an in-phase mode 
$\delta n_{\rm tot} ({\bf q}, \omega) 
\equiv \delta n_{\uparrow} ({\bf q}, \omega) + \delta n_{\downarrow} ({\bf q}, \omega)$ 
and of an out-of-phase mode $\delta n_{-} ({\bf q}, \omega) 
\equiv \delta n_{\uparrow} ({\bf q}, \omega) - \delta n_{\downarrow} ({\bf q}, \omega)$ 
exist because of the two component system. 
The hydrodynamic equations in terms of these quantities are written as 
\begin{eqnarray}
0 &=& \omega \delta n_{-} ({\bf q}, \omega) , 
\\
0 &=&\omega \delta n_{\rm tot} ({\bf q}, \omega) + \frac{2W_{2}}{3m} {\bf q}\cdot \overline{\bf v}({\bf q}, \omega), 
\\ 
0 &=&\left ( \omega - i \frac{2\eta q^{2}}{W_{2}} \right ) 
{\bf q}\cdot \overline{\bf v}({\bf q}, \omega) 
+ 
 \frac{q^{2}}{2mW_{2}} \delta E ({\bf q}, \omega) 
- \frac{g q^{2}}{2 m} \delta n_{\rm tot} ({\bf q}, \omega) , 
\\
0 &=& \left ( \omega - 4 i \kappa m^{2} T q^{2}  R \right ) \delta E ({\bf q}, \omega) 
+  \frac{2W_{4}  }{3m} {\bf q}\cdot \overline{\bf v}({\bf q}, \omega) 
- 4 i \kappa m^{2} T q^{2}  R  \frac{W_{ 2}}{W_{ 0}} 
\delta n_{\rm tot} ({\bf q}, \omega) , 
\end{eqnarray}
where $R$ is defined as $1/R \equiv 2  (W_{4} - W_{ 2}^{2}/ W_{ 0} )$, and 
the assumption of the population balanced gas $W_{n} \equiv W_{\uparrow, n}  =W_{\downarrow, n}$ 
is used. 
We note that the out-of-phase mode is decoupled from the hydrodynamic mode composed of the total density, the velocity, and the energy, in the population balanced gas.

Solving the secular equation for fluctuations 
$(\delta n_{\rm tot} ({\bf q}, \omega), {\bf q}\cdot \overline{\bf v}({\bf q}, \omega) , \delta E ({\bf q}, \omega))$, 
one obtains an equation 
\begin{align}
F_{1}(\omega) + F_{2}(\omega) = 0, 
\label{F1F2}
\end{align}
where 
\begin{align}
F_{1}(\omega) 
&\equiv 
\omega^{3} 
-
\omega 
\frac{q^{2}}{3m^{2}W_{2}} (W_{4} - gW_{2}^{2}), 
\label{seclarhydor1}
\\
F_{2}(\omega)
&\equiv 
-\omega^{2} 
\left ( 
\frac{2i\eta q^{2}}{W_{2}} + 4i\kappa m^{2} T q^{2} R 
\right )
+ \frac{4}{3} i \kappa T q^{4} R \left ( \frac{W_{4}}{W_{2}} - g W_{2} \right ). 
\label{seclarhydor2} 
\end{align} 
We omit terms of second and higher order in transport coefficients $\kappa$ and $\eta$. 
These coefficients are assumed to be small in the hydrodynamic regime.

A frequency $\omega$ of a collective excitation 
can be separated into a real part $\Omega$ and an imaginary part $\Gamma$ with being a damping rate: $\omega \equiv \Omega -i\Gamma$. 
Undamped solutions satisfy $F_{1}(\Omega) = 0$. 
Frequencies $\Omega \neq 0$ obtained from $F_{1}(\Omega) = 0$ are 
\begin{align}
\Omega_{\pm} \equiv \pm \Omega  
= 
\sqrt{
\frac{
W_{4}
-
g 
W_{2}^{2}
}
{
3W_{2}
}
}
\frac{q}{m}. 
\label{ReFirstSound}
\end{align}
A mode $\Omega = 0$ which is a thermal diffusion mode also exists, and it will be discussed in Sec. VIII. 

In the weak coupling limit at $T \rightarrow 0$, 
the frequency is given by $\Omega \approx qv_{\rm F} / \sqrt{3}$, 
where $v_{\rm F}$ is the Fermi velocity given by $v_{\rm F} \equiv (\hbar/m)(3\pi^{2}N_{\rm tot}/V)^{1/3}$. 
$N_{\rm tot}$ is the total number of particle, and $V$ is a volume. 
In the strong coupling limit at $T \rightarrow 0$, on the contrary, 
the frequency is given by 
$\Omega \approx q\sqrt{gN_{\rm tot}/(2mV)} = qv_{\rm F} \sqrt{g\rho_{\rm F}/3}$, 
where $\rho_{\rm F}$ is the density of state at the Fermi energy given 
by $\rho_{\rm F} \equiv 2m (3\pi^{2}N_{\rm tot}/V)^{1/3} / (2\pi\hbar)^{2}$. 

Damping rates and transport coefficients are assumed to be small in the hydrodynamic regime. 
The term $F_{2}(\Omega -i\Gamma)$ including transport coefficients 
can be then approximated by $F_{2}(\Omega)$, 
so that we reduce Eq. (\ref{F1F2}) as $F_{1}(\Omega -i\Gamma) + F_{2}(\Omega) = 0$. 
As a result, 
one obtains the damping rate $\Gamma_{\pm}$ as 
\begin{align}
\Gamma_{\pm}
\approx 
- i \frac{F_{2}(\Omega_{\pm}) }{\Omega_{\pm}
(\Omega_{\pm}-\Omega_{\mp})}
= 
-
\frac{\eta q^{2}}{W_{2} }
- 
\frac{\kappa  T q^{2} m^{2}}
{
W_{4}
-
g 
W_{2}^{2}
}, 
\label{ImFirstSound} 
\end{align}
where we use 
$F_{1}(\omega) = 
\omega (\omega -\Omega_{+})(\omega -\Omega_{-})$.

\section{Zero Sound}\label{zero} 
This section discusses a sound mode in the collisionless regime. 
We start with the linearized Boltzmann equation based on Eq. (\ref{Equation116}): 
\begin{align}
&
\sum\limits_{l = 0}^{\infty}
\frac{\partial f_{\sigma}^{0}}{\partial \varepsilon_{\sigma}^{0}}
\left [
\left ( 
\omega -\frac{pq}{m}\cos{\theta} 
\right )
\nu_{\sigma,l}  P_{l} (\cos{\theta})
\right ]
+ 
\frac{\partial f_{\sigma}^{0}}{\partial \varepsilon_{\sigma}^{0}}
\left (
\frac{pq}{m} \cos{\theta} 
\right )
g \langle \nu_{- \sigma, 0} \rangle
\nonumber
\\
= &
-\frac{i}{\tau}
\frac{\partial f_{\sigma}^{0}}{\partial \varepsilon_{\sigma}^{0}}
\left [\sum\limits_{l = 0}^{\infty}\nu_{\sigma} P_{l} (\cos{\theta}) 
- 
\frac{\langle \nu_{\sigma, 0} \rangle }{W_{\sigma, 0}}
- 
\frac{\langle p \nu_{\uparrow, 1} \rangle + \langle p \nu_{\downarrow, 1} \rangle}
{W_{\uparrow,2} + W_{\downarrow}} p\cos{\theta} \right ], 
\label{ZeroStart}
\end{align} 
where the simplified notations $\nu_{\sigma, l}^{m = 0}(p) \equiv \nu_{\sigma, n}(p)$ and 
$P_{l}^{m=0} (\cos{\theta}) \equiv P_{l} (\cos{\theta})$ are used. 
We assume conservation laws only for number of particles and for momentum 
as in Ref.~\cite{Khalatnikov1958}: $a_{\sigma} = \langle \nu_{\sigma, 0} \rangle / W_{\sigma, 0}$, 
$b = \sum\limits_{\sigma} \langle p\nu_{\sigma, 1} \rangle /   \sum\limits_{\sigma} W_{\sigma, 2}$, 
and $c = 0$. 

Equation (\ref{ZeroStart}) can be reduced to 
\begin{eqnarray}
&
\displaystyle{
\frac{\partial f_{\sigma}^{0}}{\partial \varepsilon_{\sigma}^{0}} 
}
\left [ 
\sum\limits_{l = 0}^{\infty} 
\nu_{\sigma,l}  P_{l} (\cos{\theta}) 
- 
g 
\displaystyle{
\frac{\cos{\theta}}{\cos{\theta} - 
\displaystyle{\frac{i \tau \omega -1}{i \tau p q/m}}
}
}
\langle \nu_{- \sigma, 0} \rangle
\right ] 
\nonumber
\\
= &
\displaystyle{
\frac{1}{i \tau}
\frac{\partial f_{\sigma}^{0}}{\partial \varepsilon_{\sigma}^{0}} 
}
\frac{1}{\displaystyle{\frac{pq}{m}}}
\displaystyle{ 
\frac{1}{\cos{\theta} - \displaystyle{\frac{i \tau \omega -1}{i \tau p q/m}}} 
}
\left [ 
\frac{\langle \nu_{\sigma, 0}\rangle }{W_{\sigma, 0}}
+ 
\frac{\langle p \nu_{\uparrow, 1}\rangle + \langle p \nu_{\downarrow, 1}\rangle   }
{W_{\uparrow, 2} + W_{\downarrow, 2}} p\cos{\theta}
\right ]. 
\end{eqnarray} 
Multiplying this equation by $1$ and $p\cos{\theta}$, and integrating over the momentum ${\bf p}$, 
we obtain the following two equations: 
\begin{align}
\left ( 
1 - g A_{0,1} - \frac{1}{i\tau q/m} \frac{A_{-1,0}}{W_{0}} 
\right ) \langle \nu_{0} \rangle
= & \frac{1}{i\tau q/m} \frac{A_{0,1}}{W_{2}} \langle p \nu_{1}\rangle, 
\label{eqKASE1}
\\
\left ( 
\frac{1}{3} - \frac{1}{i\tau q/m} \frac{A_{1,2}}{W_{2}} 
\right ) \langle p \nu_{1}\rangle
= & 
\left ( 
g A_{1,2} + \frac{1}{i\tau q/m} \frac{A_{0,1}}{W_{0}}
\right ) \langle \nu_{0}\rangle, 
\label{eqKASE2}
\end{align}
where $W_{n} \equiv W_{\uparrow, n} = W_{\downarrow, n}$. 
We consider the in-phase mode $\langle p^{n} \nu_{l}\rangle 
\equiv \langle p^{n} \nu_{\uparrow, l}\rangle + \langle p^{n} \nu_{\uparrow, l}\rangle$. 
The coefficient $A_{n,l}$ is defined as 
\begin{align}
A_{n,l} \equiv \int \frac{d{\bf p}}{(2\pi\hbar)^{3}} 
\frac{\partial f^{0}}{\partial \varepsilon^{0}} p^{n} 
\frac{\cos^{l}{\theta}}{\cos{\theta} - 
\displaystyle{\frac{i\tau\omega -1}{i\tau pq/m}}}, 
\end{align}
where the spin index in $\partial f^{0} / \partial \varepsilon^{0}$ is omitted. 

Let us consider zero sound in the low temperature regime. 
We impose the temperature dependence only to the relaxation time, 
and simply evaluate the coefficient $A_{n.l}$ as the value at $T=0$. 
From Eqs. (\ref{eqKASE1}) and (\ref{eqKASE2}), the following dispersion relation can be obtained: 
\begin{align}
1 - i \frac{s''}{s} + W^{0}(s) 
\left [ g\rho_{\rm F} + i \frac{s''}{s} + i 3 s'' \left ( s - i s'' \right ) \right ] = 0. 
\label{eqKA}
\end{align}
The function $W^{0} (s)$ is the Lindhard function given by 
\begin{align}
W^{0} (s) = 1 - \frac{s}{2} \ln{\left | \frac{s + 1}{s - 1} \right |}. 
\end{align}
$s$ and $s''$ are defined as $s \equiv (i\tau \omega -1)/(i \tau v_{\rm F} q)$, 
and $s'' \equiv 1/(\tau v_{\rm F}q)$, respectively. 
The dispersion relation (\ref{eqKA}) is first derived by Khalatnikov and Abrikosov~\cite{Khalatnikov1958}. 
The notations used here follows in Ref.~\cite{Larionov1999}.

The frequency of the collective mode in collisionless limit $\tau\omega \gg 1$ is given by 
\begin{align}
1 + g\rho_{\rm F}\left ( 1- \frac{\Omega}{2v_{\rm F}q} 
\ln{\left | \frac{\Omega+v_{\rm F}q}{\Omega- v_{\rm F}q}\right |} \right ) = 0, 
\label{ReZeroSound}
\end{align}
where we omit $s''$ in the dispersion relation (\ref{eqKA}) because it is small. 
This result can be also obtained by the random phase approximation discussed in Appendix~\ref{RPA}. 
The frequency is given by 
$\omega = q v_{\rm F} \{ 1+2\exp{[-1/(g\rho_{\rm F})]} \}$ in the weak coupling limit $g\rho_{\rm F} \ll 1$; 
the frequency is given by $\omega = q v_{\rm F} \sqrt{g\rho_{\rm F}/3}$ in the strong coupling limit $g\rho_{\rm F} \gg 1$.  

The damping rate $\Gamma = - {\rm Im}(\omega)$ in the collisionless limit $\tau\omega \gg 1$ 
is obtained by the following way.  
We expand the dispersion relation (\ref{eqKA}) to first order in $s''$ and $\Gamma$ 
since these are small. 
As a result, we obtain the damping rate $\Gamma$ as 
\begin{align}
\Gamma = \frac{1}{\tau} \left [ 1- \frac{(g\rho_{\rm F} + 1 + 3 s_{0}^{2})(s_{0}^{2} - 1)}
{g\rho_{\rm F} (g \rho_{\rm F} + 1 - s_{0}^{2})} \right ], 
\label{ImZeroSound}
\end{align}
where $s_{0} \equiv \Omega/v_{\rm F} q$. 
This is seen in Ref.~\cite{Larionov1999}. 

The frequency and the damping rate in the low frequency regime $\omega \tau \ll 1$ 
can be also evaluated based on the dispersion relation (\ref{eqKA}). 
The Lindhard function $W^{0}(s)$ is approximated as $W^{0}(s) \simeq - 1/3s^{2} - 1/5s^{4}$ in this regime. 
When we consider the dispersion relation (\ref{eqKA}) with the first order of the relaxation time $\tau$, 
we obtain 
\begin{align}
\left ( \frac{\omega}{v_{\rm F}q} \right )^{2}  = \frac{1}{3} (g \rho_{\rm F}+1) -\frac{4}{15} i \omega \tau. 
\label{collisionalKA}
\end{align}
As a result, the frequency of the collective mode is given by $\Omega = q v_{\rm F}\sqrt{(1+g\rho_{\rm F})/3}$. 
One obtains $\Omega \simeq q v_{\rm F}\sqrt{g\rho_{\rm F}/3}$ 
in the strong coupling case $g\rho_{\rm F} \gg 1$. 
This corresponds to the frequency of the first sound at $T=0$ in Eq. (\ref{ReFirstSound}) in the strong coupling limit. 
The damping rate can be approximately evaluated 
as $\Gamma \simeq 2 \tau (v_{\rm F}q)^{2}/15$ from Eq. (\ref{collisionalKA}). 
This damping rate is consistent with the damping rate of the first sound in Eq. (\ref{ImFirstSound}), 
if we impose the temperature dependence in Eq. (\ref{ImFirstSound}) 
only to the relaxation time, 
and evaluate other quantities in Eq. (\ref{ImFirstSound}) as the value at $T = 0$. 
The viscous term $-\eta q^{2}/W_{2}$ alone contributes this damping rate.

\section{Results : Sound mode from $\omega \tau \ll 1$ to $\omega \tau \gg 1$}\label{results}

This section presents the results obtained by solving the moment equation (\ref{MomentEquation}). 
We focus on the crossover from zero sound to first sound. 

The collisionless regime $\omega \tau \gg 1$ and the collisional regime $\omega \tau \ll 1$ 
can be realized by controlling the temperature $T$. 
In the high temperature regime, atoms are colliding with each other frequently, so that the hydrodynamic regime $\omega \tau \ll 1$ is achieved. 
In the low temperature regime, on the contrary, 
the Pauli blocking makes the phase volume where the atoms are scattered restricted, 
and hence the collisionless regime $\omega \tau \gg 1$ is realized. 

The coupling constant $\alpha = gN_{\rm tot}/(V\varepsilon_{\rm F})$ also plays a role determining 
the collisionless and collisional regimes, 
where $\varepsilon_{\rm F}$ is the Fermi energy 
given by $\varepsilon_{\rm F} \equiv [\hbar^{2}/(2m)] (3\pi^{2}N_{\rm tot}/V)^{2/3} $.
The mean-field potential is proportional to the coupling constant $\alpha$, 
while the collision integral (or the relaxation rate) is proportional to $\alpha^{2}$. 
The collisionless regime $\omega \tau \gg 1$ could be realized in the weak coupling regime. 
In the strong coupling regime, on the contrary, the collision term (or the relaxation rate) is dominant compared with the mean-field potential, 
so that one would be in the collisional regime. 
From these points of view, we study the sound mode from $\omega \tau \ll 1$ to $\omega \tau \gg 1$ 
as a function of the temperature $T$ and of the coupling constant $\alpha$. 

\begin{figure} 
\begin{center}
\includegraphics[width=13cm,height=10cm,keepaspectratio,clip]{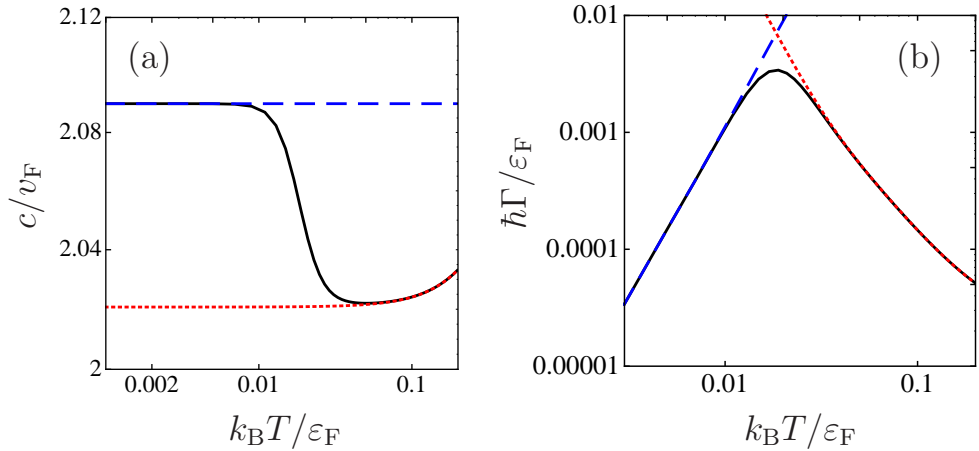}
\end{center}
\caption{
Frequency and damping rate of collective excitation as a function of temperature. 
Figure (a) shows the phase velocity. Figure (b) shows the damping rate. 
Solid lines correspond to an eigenmode obtained from the moment equation. 
Dashed lines in (a) and (b) represent the phase velocity and the damping rate of zero sound, respectively. 
Dotted lines in (a) and (b) represent those of first sound. 
The coupling constant $\alpha = 15$ is used. 
}
\label{Fig1}
\end{figure}

In Fig.~\ref{Fig1}, 
eigenvalue $\omega$ of the collective excitation is plotted as a function of $T$. 
We show a strong coupling case  $\alpha = 15$. 
We chose the wavenumber $q = 0.05k_{\rm F}$. 
We take moments up to $l = 30$ and $n = 30$ in this calculation, 
although less moments, for example up to $l = 10$ and $n = 10$, reproduces the same result. 

Figure~\ref{Fig1} (a) shows the phase velocity $c$ defined by $c \equiv \Omega/q$ 
where $\Omega \equiv {\rm Re}(\omega )$. 
Figure~\ref{Fig1} (b) 
shows damping rate $\Gamma$ given by $\Gamma = - {\rm Im}(\omega )$. 
Solid lines in Figs.~\ref{Fig1} (a) and (b) represent the phase velocity and the damping rate 
obtained from the moment equation (\ref{MomentEquation}). 
Dashed lines in Figs.~\ref{Fig1} (a) and (b) represent those of zero sound 
obtained from Eq. (\ref{ReZeroSound}) and given in Eq. (\ref{ImZeroSound}), respectively. 
Dotted lines in Figs.~\ref{Fig1} (a) and (b) represent those of first sound 
given in Eq. (\ref{ReFirstSound}) and Eq. (\ref{ImFirstSound}). 
Solutions of the moment method coincide with asymptotic solutions in two limiting regimes: 
collisionless and hydrodynamic regimes.  
Note that the moment equations show the crossover between the zero and first sound modes 
as well as the temperature dependence of first sound. 
Corresponding behavior of our result is also seen in the experimental result in Ref.~\cite{Abel1966}, 
which reported temperature dependence of the sound velocity and the amplitude attenuation coefficient 
of liquid $^{3}$He.

In the collisional hydrodynamic regime $\omega\tau \ll 1$, 
the dispersion relation in Eq. (\ref{eqKA}) first derived by Khalatnikov and Abrikosov~\cite{Khalatnikov1958} cannot reproduce our results correctly, 
because the temperature dependence is imposed only to the relaxation rate (see also Eq. (\ref{collisionalKA})). 
In Eq. (\ref{ReFirstSound}), $W_{4}$, which is proportional to the pressure, strongly depends on temperature, 
and this brings temperature dependence of the sound velocity of first sound. 
($W_{2}$, which is proportional to the density, does not have the temperature dependence under a fixed volume.) 

As for the damping rate, 
the dispersion relation in Eq. (\ref{collisionalKA}) does not involve the contribution of the thermal conductivity. 
Even if we neglect the second term in Eq. (\ref{ImFirstSound}), 
the dispersion relation in Eq. (\ref{collisionalKA}) still does not reproduce our result, 
although the difference is quite small. 
The difference also comes from the temperature dependence of the pressure in the term $W_{4}$, 
which is not involved in Eq. (\ref{eqKA}). 
It is, nevertheless, remarkable that the dispersion relation (\ref{eqKA}) first derived by Khalatnikov and Abrikosov~\cite{Khalatnikov1958} 
excellently grasps the sound velocities and damping rates in both collisionless and hydrodynamic regimes.

\begin{figure}
\begin{center}
\includegraphics[width=13cm,height=10cm,keepaspectratio,clip]{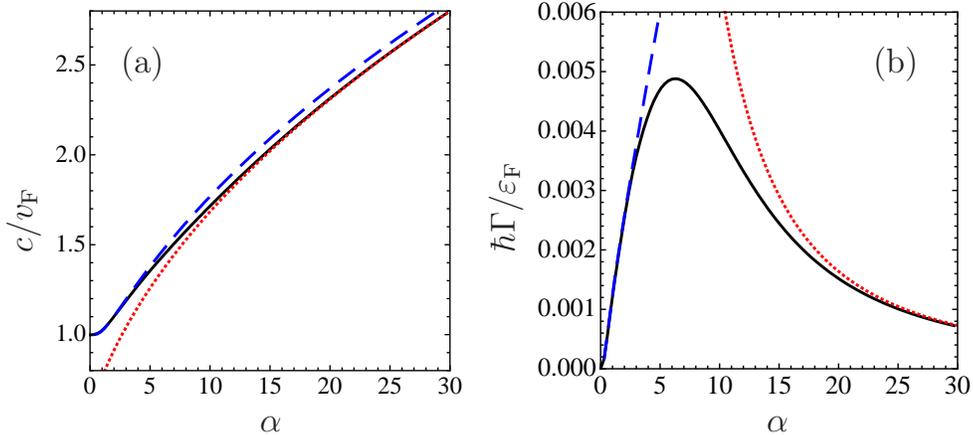}
\end{center}
\caption{
Phase velocity in (a) and damping rate in (b) are shown 
as a function of the coupling constant $\alpha = gN_{\rm tot}/V\varepsilon_{\rm F}$, 
fixing the temperature at $T = 0.025\varepsilon_{\rm F}$.  
Solid lines in (a) and (b) show the results obtained by the moment method. 
The dashed lines in (a) and (b) are  the phase velocity and the damping rate of zero sound, respectively. 
The dotted lines in (a) and (b) represent those of first sound, respectively. 
}
\label{Fig2}
\end{figure}

In turn, 
we plot the phase velocity $c$ and the damping rate $\Gamma$ of the collective mode 
as a function of the coupling constant $\alpha$ in Figs. \ref{Fig2} (a) and (b). 
We show the low temperature case $k_{\rm B}T = 0.025\varepsilon_{\rm F}$.   
Again, we choose the wavenumber $q = 0.05k_{\rm F}$,  
and take moments up to $l = 30$, and $n = 30$.   
Solid lines in Figs. \ref{Fig2} (a) and (b) 
show the phase velocity and the damping rate obtained from the moment equation (\ref{MomentEquation}). 
The dashed lines and the dotted lines in Figs. \ref{Fig2} (a) and (b) 
show the velocity and damping of zero sound (given in Eqs. (\ref{ReZeroSound}) and (\ref{ImZeroSound})) 
and first sound (given in Eqs. (\ref{ReFirstSound}) and (\ref{ImFirstSound})), respectively. 
The crossover from zero sound to first sound can be clearly seen in this figure. 

From Fig. \ref{Fig2} (a), one can see that the phase velocity of zero sound is close to that of first sound 
in the strong coupling regime. 
This is due to the fact that the phase velocity $c$ of zero sound is given by the same formula as that 
of first sound $c \approx v_{\rm F}\sqrt{g\rho_{\rm F}/3}$ in the strong coupling limit at $T = 0$. 
Note that the mechanisms of sound propagation are quite different in two regimes.

One could change the coupling constant $\alpha = gN_{\rm tot}/(V\varepsilon_{\rm F})$ 
by controlling a density $N_{\rm tot}/V$, 
or an interaction strength $g$ through the Feshbach resonance. 
The Fermi energy is also a function of the density, 
{\it i.e.}, $\varepsilon_{\rm F} \propto (N_{\rm tot}/V)^{2/3}$, 
and hence the coupling constant $\alpha$ has a density dependence: $\alpha \propto (N_{\rm tot}/V)^{1/3}$.

\section{discussion}\label{discussion} 
In this section, we discuss physical implication of results obtained from the moment equation. 
First, we discuss a hydrodynamic mode other than the sound mode. 
Second, excitation spectrum of the particle-hole continuum obtained from the moment equation is discussed. 
Third, we discuss the sound mode in a weak coupling case is made. 
Forth, other issues and future problems are discussed. 

\subsection{Thermal Diffusion Mode}

In discussing the collective mode in the hydrodynamic regime, 
there usually exist five modes, corresponding to the particle number, the velocity and the energy. 
Two modes are the first sound modes $\pm \Omega - i \Gamma$, 
related to the particle number and the velocity of a certain direction, discussed in Sec.~\ref{first}. 
Other two modes are shear modes $\Gamma_{\eta}$ 
related to the velocity of remaining two directions. 
The other is the thermal diffusion mode $\Gamma_{\kappa}$ related to the energy. 
Note that the shear modes and the thermal diffusion mode are purely damping modes.

In the present paper, we assume that vectors ${\bf b}$ and ${\bf q}$ are parallel each other 
as treated in Sec.~\ref{moment} and Sec.~\ref{first}. 
This means that velocity of the fluid ${\bf v}$ is assumed to be parallel 
to the wavenumber vector of the collective mode ${\bf q}$, 
and hence two shear modes are neglected. 
In this subsection, we discuss the thermal diffusion mode. 

In Sec.~\ref{first}, we noted that a collective mode $\Omega = 0$ exists. 
Assume that the result is written as $\omega = - i \Gamma$, 
and consider the term up to the first order in damping rate and transport coefficients 
in Eqs. (\ref{seclarhydor1}) and (\ref{seclarhydor2}). 
Setting $F_{1}(-i\Gamma) + F_{2}(0) = 0$,  
we obtain damping rate of the thermal diffusion mode 
\begin{align}
\Gamma_{\kappa} 
= - \frac{2 \kappa T m^{2} q^{2} W_{2}^{2} (1-gW_{0})}{(W_{4}-gW_{2}^{2})(W_{4}W_{0}-W_{2}^{2})}. 
\label{HeatDiffusion}
\end{align} 

\begin{figure}[htbp]
\begin{center}
\includegraphics[width=13cm,height=10cm,keepaspectratio,clip]{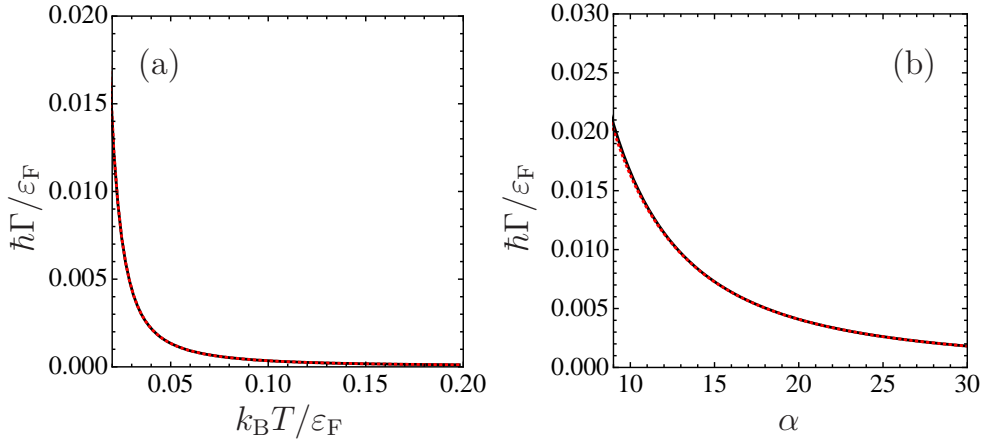}
\end{center}
\caption{
Thermal diffusion modes are plotted 
as a function of temperature in (a), 
and of the coupling constant in (b). 
Solid lines in (a) and (b) are the results obtained by the moment method. 
The dotted lines represent thermal diffusion modes given in Eq. (\ref{HeatDiffusion}). 
}
\label{Fig3}
\end{figure}

Since the moment method provides the general solution of the linearized Boltzmann equation, 
the thermal diffusion mode should be also included. 
In Fig.~\ref{Fig3}, the damping rates corresponding to the thermal diffusion mode are plotted. 
In this calculation, we take moments up to $l = 30$ and $n = 30$, and chose the wavenumber $q = 0.05k_{\rm F}$. 
The solid lines are the damping rates obtained from  the moment equation (\ref{MomentEquation}). 
The dotted lines are the damping rates given in Eq. (\ref{HeatDiffusion}). 
In Fig. 4 (a), the damping rate of the thermal diffusion mode versus temperature is shown 
for the coupling constant $\alpha=15$. 
In Fig. 4 (b), the damping rate versus the coupling constant $\alpha$ is shown 
for the temperature $k_{\rm B}T = 0.025\varepsilon_{\rm F}$. 
Parameters in Figs. (a) and (b) are the same as in Fig.~\ref{Fig1} and Fig.~\ref{Fig2}, respectively. 
We confirm that the present moment method provides the thermal diffusion mode. 

\subsection{Particle-Hole Continuum}

In discussing zero sound, 
one often uses the random phase approximation. 
The usual random phase approximation (see Appendix~\ref{RPA}) gives 
excitation spectra in the particle-hole continuum as well as a collective mode. 
We discuss excitation spectra at $T=0$ obtained by solving the moment equation from this point of view. 

\begin{figure}[htbp]
\begin{center}
\includegraphics[width=8cm,height=8cm,keepaspectratio,clip]{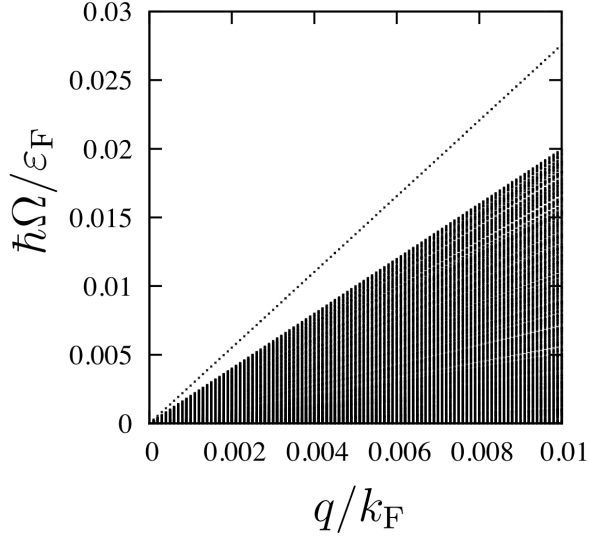}
\end{center}
\caption{
Eigenvalues $\Omega = {\rm Re}(\omega) $ as a function of the wavelength $q$ at $T=0$, 
obtained from the moment equation. }
\label{Fig4}
\end{figure} 

In Fig.~\ref{Fig4}, 
we plot real part of eigenvalues obtained from the moment equation (\ref{MomentEquation}) 
as a function of the wavelength $q$ at $T = 0$. 
The real part of these frequencies is symmetric with respect to the $q$-axis, 
so that we show only the region $\Omega > 0$ in Fig~\ref{Fig4}. 
We take moments up to $l = 30$ and $n =30$ in the numerical calculation. 
The coupling constant $\alpha \equiv gN_{\rm tot}/(V\varepsilon_{\rm F}) = 5$ is used. 
From Fig.~\ref{Fig4}, one finds that real parts of eigenvalues in the moment equation 
also include excitation spectra corresponding to the particle-hole continuum as well as the collective excitation. 
The gradient of the edge of the particle-hole continuum excitation in this figure is seen to be $2$ 
in our dimensionless units, which corresponds to 
$\Omega = v_{\rm F}q$ in the real physical units. 

In the usual random phase approximation, 
the denominator of the response function is given by 
$\omega + \varepsilon_{{\bf p}} - \varepsilon_{{\bf p} + {\bf q}}$, 
and hence spectrum includes $\omega = {\bf p}\cdot{\bf q}/m + q^{2}/(2m)$~\cite{NegeleBOOK}. 
This feature brings the phonon excitation $\omega \propto q$ 
at the long-wavelength regime $q/k_{\rm F} \ll 1$, 
and the parabolic excitation $\omega \propto q^{2}$ at $q/k_{\rm F} \gg 1$, 
where $k_{\rm F}$ is the Fermi wavenumber. 
Solution of the semiclassical Boltzmann equation only involves the denominator $\omega - {\bf p}\cdot{\bf q}/m$, 
as seen in Eq (\ref{RPA-2}),
and hence our calculation can reproduce only the phonon regime: $\Omega \propto q$.

We presented discussion of the particle-hole continuum, 
but we remark some issues shown in~\ref{Remarks}. 

\subsection{On the Weak Coupling Case}

The phase velocity of zero sound is always larger than the Fermi velocity when $g > 0$. 
The phase velocity of first sound, however, could be less than the Fermi velocity in the weak coupling case 
and at low temperatures. 
In such cases, 
the spectrum of the collective excitation is not necessarily pushed up above the particle-hole continuum. 
We discuss the results in such a weak coupling case.

We calculate phase velocities $c \equiv {\rm Re}(\omega )/q$ as a function of the temperature 
in $\alpha = 1$. 
We chose $q = 0.01k_{\rm F}$. 
In the calculation, we take moments up to $l = 11$ and $n = 11$. 
A reason of truncation at the moderate moments is that it allows us to clearly see transitions of the each eigenvalue. 
At $T \simeq 0$, zero sound is seen as a separate eigenvalue where $c \sim v_{\rm F}$. 
At finite temperatures, 
we confirmed that the eigenvalue of the collective excitation is buried in the particle-hole continuum. 
In this calculation, we also found that the spectra of zero sound and of first sound are not continuous 
in contrast with the strong coupling case. 
This result suggests that a collective mode in a weakly coupling system has a different feature 
from that in a strongly coupling system. 
It is unclear that how the collective mode behaves in the weakly coupling system in the crossover regime, 
even if more moments are taken. 
In a separate paper, we will calculate the dynamic structure factor of a normal Fermi system at finite temperatures, 
and discuss this problem~\cite{Watabe2009}.

\subsection{Remarks and Future Problems}\label{Remarks}

Before closing this section, 
we make some remarks on results of the present method and propose future problems. 

We first note some issues on the present method. 
Even if we set the relaxation rate $1/\tau$ to be zero, 
the coefficient matrix of the moment equation (\ref{CoefficientMatrix}) 
is not symmetric, although the matrix elements are real. 
The eigenvalues are thus complex, in general. 
Those damping rates, namely, the imaginary parts of the resulting eigenvalues, 
range from $-v_{\rm F}q$ to $v_{\rm F}q$ at $T=0$, 
and hence the present moment method could not reproduce the results of the random phase approximation perfectly. 
We note that those damping rates increase 
as the temperature $T$ or the coupling constant $\alpha$ increases. 
In addition, there exists an additional purely damped mode, 
which is not included in the sound mode or the thermal diffusion mode. 
This mode does not belong to the complex eigenvalues discussed above either. 
The damping rate of this mode could be negative at certain temperatures and certain coupling constant. 
Albeit the present moment method involves the issues mentioned above, 
we insist that the present method offers very intriguing studies on the collective mode over a wide range of parameters. 

In turn, we shall discuss the future problem from the physical point of view.  

The excitation spectra in the weakly interacting system is complicated as discussed in Sec.~\ref{discussion} C. 
The spectrum of the collective excitation buried in the particle-hole continuum in the crossover regime. 
The collective mode and the single particle excitations are strongly related in this crossover regime, 
and hence the effect of Landau damping could be important. 
One issue is how the feature of the collective mode remains or disappears in this regime. 
In a separate article, we will study this problem by the dynamic structure factor~\cite{Watabe2009}. 
The effect of the Landau damping would be seen in the peak width of the dynamic structure factor.

It is also interesting to solve the equation derived by Brooker and Sykes~\cite{Brooker1970}. 
The equations in Ref.~\cite{Brooker1970} involves additional equation in order to close the moment equation. 
Ref.~\cite{Brooker1970}, in addition, introduces different relaxation times for different moments. 
Such a treatment is complicated compared with our formulation, 
and it is not obvious how the additional equation affects our result. 

%We also note that one can also study transverse zero sound, and the shear mode, extending this method. 

The Landau's Fermi liquid theory focuses on the low temperature property, 
since this theory is based on an idea 
that a lifetime of quasi-particles are sufficiently long at very low temperatures. 
For this circumstance, the crossover between the zero and first sound modes 
has been studied theoretically only within the low temperature approximation. 
In ultracold Fermi gases, 
the real-particle picture is also important in both a classical gas regime 
and a weakly interacting Fermi system. 
Our formulation allows one to describe such a system. 
The temperature $T$, the density $N_{\rm tot}/V$ 
and also the interaction strength $g$ are controllable 
with recent techniques in ultracold atomic gases. 
We expect that behaviors of the collective mode shown in the present paper 
would be observed in the experiments of ultracold Fermi gases.

We now comment on the application of the present work 
to a strongly interacting Fermi gas near the unitarity limit. 
At sufficiently high temperatures of the Maxwell-Boltzmann regime, 
real particles are important, and thus a gas is described by the Boltzmann equation 
with an energy dependent cross section~\cite{Massignan2005}. 
In contrast, in the low temperature regime above the superfluid transition temperature, 
the system may be described by the Landau's Fermi liquid theory for quasiparticles. 
One expects a crossover from quasiparticle picture to real particle picture with increasing temperature, 
which may be studied by the moment method developed in the present paper. 
%The Landau's Fermi liquid theory is based on the assumption 
%that the system is described in terms of a gas of long-living quasiparticles. 
Explicit determination of the range of $s$-wave scattering length as well as of the temperature, 
where the kinetic equation analysis based on the long-living quasiparticle picture is valid, 
will require many-body calculation for a Fermi gas near the unitarity limit~\cite{Bruun2004}.

\section{summary and conclusion}\label{summary}

The moment method is suitable for describing the collective mode from collisionless to collisional regimes 
with only a relaxation time approximation. 
We solved the linearized Boltzmann equation for a normal Fermi system using this method, 
and obtained the general solution. 
We discussed the crossover between the zero and first sound modes  
as a function of the temperature and the coupling constant. 
We found that 
an eigenfrequency of a collective mode obtained from the moment equations 
reproduces the sound velocity and the damping rate in the crossover regime as well as both collisionless and collisional limiting regimes. 

Through the analysis of the moment equation, 
we found that the moment method provides the thermal diffusion mode. 
We also discussed the excitation spectra of the particle-hole continuum, 
and the sound mode in a weak coupling case. 
We finally made remarks on the present method and proposed future problems.

\section{acknowledgment}
We thank S. Konabe, T. Miyakawa, and C. Tachibana 
for helpful discussions. 
S. W. thanks Y. Kato for valuable comments. 
S. W. acknowledges support from 
the Fujyu-kai Foundation, 21st Century COE Program at University of Tokyo, 
GCOE for Phys. Sci. Frontier, MEXT, Japan, and Grant-in-Aid for JSPS Fellows (217751). 
T. N. was supported by Grant-in-Aid for Scientific Research from JSPS. 

\appendix

\setcounter{section}{0}

\section{Chapman-Enskog Method and Transport Coefficients}\label{chapman}

In this section, we give a derivation of transport coefficients in a degenerate Fermi gas 
based on the Chapman-Enskog method. 
The result in this section will be used to evaluate the relaxation time in the next section. 

The transport coefficient in the Landau's Fermi liquid was first calculated 
by Abrikosov and Khalatnikov~\cite{Abrikosov1957}. 
Afterwards, it was analyzed in several papers~\cite{Hone1960, Dy1968, Dy1969, Sykes1970}.  
The Chapman-Enskog method was first generalized to quantum gases 
by Uehling and Uhlenbeck~\cite{Uehling1933,Uehling1934}. 
In this Appendix, the analysis is based on Ref.~\cite{Nikuni1998}. 

Following the standard procedure, 
we define the following hydrodynamic physical quantities: 
\begin{eqnarray}
\mbox{density :}
&&
n_{\sigma} (\coo) \equiv \intmom f_{\sigma}({\bf p}, \coo) ,
\label{density} 
\\
\mbox{ total density :} 
&&
 n_{\rm tot} (\coo) \equiv \sum\limits_{\sigma}n_{\sigma} (\coo) , 
\label{totaldensity} 
\\
\mbox{ velocity :}
&&
n_{\sigma}(\coo) {\bf v}_{\sigma}(\coo) \equiv \intmom \frac{\bf p}{m} f_{\sigma} ({\bf p}, \coo) ,
\label{velocity} 
\\
\mbox{ pressure tensor:} 
&&
P_{\mu\nu}(\coo) 
\equiv 
\sum\limits_{\sigma}
P_{\sigma, \mu\nu}(\coo) 
\nonumber
\\
&&
\qquad \qquad
\equiv 
\sum\limits_{\sigma}
m
\intmom 
\left [ 
\frac{p_{\mu}}{m} - v_{\sigma,\mu}(\coo) 
\right ]
\left [ 
\frac{p_{\nu}}{m} - v_{\sigma,\nu}(\coo)
\right ] 
f_{\sigma}({\bf p}, \coo),
\nonumber
\\
\label{pressure} 
\\
\mbox{ energy density :} 
&&
E (\coo)
\equiv 
\sum\limits_{\sigma}
E_{\sigma} (\coo)
\nonumber
\\
&&
\qquad \quad
\equiv 
\sum\limits_{\sigma}
\intmom 
\frac{1}{2m}
\left [
{\bf p} - m {\bf v}_{\sigma}(\coo)
\right ]^{2} 
f_{\sigma} ({\bf p}, \coo), 
\label{energydesity} 
\\
\mbox{ heat current :}
&&
{\bf Q} (\coo) 
\equiv 
\sum\limits_{\sigma}
 {\bf Q}_{\sigma} (\coo) 
\nonumber
\\
&&
\qquad \quad
\equiv 
\sum\limits_{\sigma}
\intmom 
\frac{1}{2m}
\left [
{\bf p} - m {\bf v}_{\sigma}(\coo)
\right ]^{2} 
\left [
\frac{\bf p}{m} - {\bf v}_{\sigma}(\coo) 
\right ]
f_{\sigma} ({\bf p}, \coo), 
\nonumber
\\
\label{heatcurrent} 
\\
\mbox{ rate-of-strain tensor :}
&&
D_{\sigma, \mu\nu} (\coo) 
\equiv 
\frac{1}{2}
\left [  
\frac{\partial {v}_{\sigma,\mu}(\coo)}{\partial x_{\nu}} 
+ 
\frac{\partial {v}_{\sigma,\nu}(\coo)}{\partial x_{\mu}} 
\right ]. 
\label{rate-of-straintensor}
\end{eqnarray}
Indexes $\mu$ and $\nu$ are Cartesian components. 

We assume that 
local velocities of two components are the same
$
v_{\mu}(\coo) 
\equiv 
v_{\uparrow, \mu}(\coo) 
= 
v_{\downarrow, \mu}(\coo)  
$. 
This means that a rate-of-strain tensor is the same for two components, 
and hence we define $D_{\mu\nu} (\coo) \equiv D_{\sigma, \mu\nu} (\coo) $. 
With the above quantities, 
generalized hydrodynamic equations are given by 
\begin{eqnarray}
&&
\frac{\partial }{\partial t}
n_{\sigma}(\coo)
+ 
\nabla_{\bf r} \cdot 
\left [
n_{\sigma}(\coo)
{\bf v}(\coo)
\right ]
=0, 
\label{hydro-density}
\\
&&
m 
n_{\rm tot}(\coo)
\left [
\frac{\partial}{\partial t}
+ 
v_{\nu}(\coo)\frac{\partial}{\partial x_{\nu}}
\right ]
v_{\mu}(\coo)
=
- 
\frac{\partial}{\partial x_{\nu}}
P_{\mu\nu}(\coo)
-
\frac{\partial }{\partial x_{\mu}} 
\left [
g n_{\uparrow} (\coo) n_{\downarrow} (\coo)
\right ], 
\label{hydro-momentum}
\\
&&
\frac{\partial}{\partial t} 
E(\coo) 
+ 
\nabla_{\bf r} {\bf Q} (\coo) 
+ 
\nabla_{\bf r} 
\left [
E(\coo) {\bf v}(\coo)
\right ] 
+ 
\sum\limits_{\mu\nu}
D_{\mu\nu}(\coo)P_{\mu\nu}(\coo)
= 0. 
\label{hydro-energy}
\end{eqnarray} 
These hydrodynamic equations are obtained 
by multiplying Eq. (\ref{Boltzmann}) by $1$, $\bf p$ and $p^{2}$ 
and integrating over ${\bf p}$. 
The collision integral in Eq. (\ref{Boltzmann}) vanishes owing to the conservation law. 

In the collision-dominated regime, the first approximation to the distribution function is 
the local equilibrium distribution $\tilde{f}_{\sigma} ({\bf p}, \coo)$. 
In local equilibrium, the hydrodynamic quantities are given by 
\begin{align}
\tilde{n}_{\sigma}(\coo) 
&= \frac{1}{\Lambda^{3}(\coo)}
{\mathcal F}_{3/2}(z_{\sigma}(\coo)), 
\label{local-density}
\\ 
\tilde{P}_{\mu\nu}(\coo) 
&= \delta_{\mu\nu} \tilde{P}(\coo) 
= 
\delta_{\mu\nu} 
\sum\limits_{\sigma}
\frac{k_{\rm B}T(\coo)}{\Lambda^{3} (\coo)}
{\mathcal F}_{5/2}(z_{\sigma}(\coo)), 
\label{local-pressure}
\\
\tilde{P}(\coo) &= \frac{2}{3} \tilde{E}(\coo), 
\label{local-pressure-energy}
\\
\tilde{\bf Q}(\coo) &= 0, 
\label{local-heat-current}
\end{align}
where $\Lambda (\coo) $ is the local thermal de Broglie wavelength: 
\begin{align}
\Lambda (\coo) 
\equiv  
\left [ 
\frac{2\pi\hbar^{2}}{mk_{\rm B}T(\coo)}
\right ]^{1/2}. 
\label{local-thermal-deBroglie}
\end{align}
${\mathcal F}_{n}(z_{\sigma}(\coo))$ is the Fermi function~\cite{Williams2004} given by 
\begin{align}
{\mathcal F}_{n}(z_{\sigma})
 = 
\frac{1}{
\Gamma\left ( n \right )
} 
\int dx 
\frac{x^{n-1}}{\exp{(x)}z_{\sigma}^{-1}+1}, 
\label{Fermi-function}
\end{align}
where $\Gamma (n)$ is the Gamma function. 
With the above quantities, 
hydrodynamic equations in local equilibrium are given by 
\begin{eqnarray}
&&
\frac{\partial }{\partial t}
\tilde{n}_{\sigma}(\coo) 
+ 
\nabla_{\bf r}
\left [
\tilde{n}_{\sigma}(\coo)\overline{\bf v}(\coo)
\right ] 
= 0, 
\label{local-density-equation}
\\
&&
m \tilde{n}_{\rm tot}(\coo) 
\left[ 
\frac{\partial }{\partial t} 
+ 
\overline{\bf v}(\coo) \cdot \nabla_{\bf r}
\right ]
\overline{\bf v}(\coo)
= 
- 
\nabla_{\bf r}\cdot \tilde{P}(\coo) 
- 
\nabla 
\left [
g \tilde{n}_{\uparrow}(\coo) \tilde{n}_{\downarrow}(\coo)
\right ],  
\label{local-momentum-equation}
\\
&&
\frac{\partial}{\partial t}\tilde{E}(\coo) 
+ 
\frac{5}{3}
\nabla_{\bf r} \left [ \tilde{E}(\coo) \overline{\bf v} (\coo) \right ] 
= 
\overline{\bf v}(\coo)\cdot 
\left [ \nabla_{\bf r}\tilde{P}(\coo) \right ].  
\label{local-energy-equation}
\end{eqnarray}

In order to treat departure from local equilibrium, 
we introduce the following form of the distribution function: 
\begin{align}
f_{\sigma}({\bf p}, \coo)
&= 
\tilde{f}_{\sigma}({\bf p}, \coo)
+ 
\tilde{f}_{\sigma}({\bf p}, \coo)
\left [
1- \tilde{f}_{\sigma}({\bf p}, \coo)
\right ] 
\Psi_{\sigma}(\coo). 
\label{discrepancy-from-localeq}
\end{align}
Since the number of particle, the total momentum, and the total energy is conserved, 
the following three constraints are imposed; 
\begin{align}
\intmom \tilde{f}_{\sigma}({\bf p}, \coo) 
\left [ 1 - \tilde{f}_{\sigma}({\bf p}, \coo)  \right ]
\Psi_{\sigma} ({\bf p}, \coo) = 0, 
\label{constraint1}
\\
\sum\limits_{\sigma}
\intmom 
p_{\mu}
\tilde{f}_{\sigma}({\bf p}, \coo) 
\left [ 1 - \tilde{f}_{\sigma}({\bf p}, \coo)  \right ]
\Psi_{\sigma} ({\bf p}, \coo) = 0, 
\label{constraint2}
\\
\sum\limits_{\sigma}
\intmom 
p^{2}
\tilde{f}_{\sigma}({\bf p}, \coo) 
\left [ 1 - \tilde{f}_{\sigma}({\bf p}, \coo)  \right ]
\Psi_{\sigma} ({\bf p}, \coo) = 0. 
\label{constraint3}
\end{align} 
The local equilibrium distribution (\ref{local-distribution}) satisfies the detail balance of the scattering 
$
\left [ 1- \tilde{f}_{\sigma}(1) \right]
\left [ 1- \tilde{f}_{- \sigma}(2) \right] 
\tilde{f}_{- \sigma}(3)
\tilde{f}_{\sigma}(4) 
= 
\tilde{f}_{\sigma}(1)
\tilde{f}_{- \sigma}(2)
\left [ 1- \tilde{f}_{- \sigma}(3) \right]
\left [ 1- \tilde{f}_{\sigma}(4) \right] 
$. 
With a use of this relation, the collision integral in the right hand side of the Boltzmann equation reduces to 
\begin{align}
{\mathcal I}_{\rm coll} [f_{\sigma}(1)] 
\equiv & 
\hat{L}_{\sigma} [\Psi_{\sigma}(1)] 
\nonumber
\\
\equiv& 
\frac{2\pi g^{2}}{\hbar}
\int \frac{d{\bf p}_{2}}{(2\pi\hbar)^{3}}
\int \frac{d{\bf p}_{3}}{(2\pi\hbar)^{3}}
\int d{\bf p}_{4}
\delta ({\bf p}_{1} + {\bf p}_{2} - {\bf p}_{3} - {\bf p}_{4})
\delta 
\left (
\frac{p_{1}^{2}}{2m} + \frac{p_{2}^{2}}{2m} 
- \frac{p_{3}^{2}}{2m} - \frac{p_{4}^{2}}{2m}
\right ) 
\nonumber
\\
& 
\times 
\left [ 1- \tilde{f}_{\sigma}(1) \right]
\left [ 1- \tilde{f}_{- \sigma}(2) \right] 
\tilde{f}_{- \sigma}(3)
\tilde{f}_{\sigma}(4) 
\left [ 
\Psi_{\sigma}(4) + \Psi_{-\sigma}(3)
- \Psi_{- \sigma}(2) - \Psi_{\sigma}(1)
\right ]. 
\label{collision-discrepancy-from-localeq}
\end{align} 
It is useful to introduce dimensionless momentum variable 
$\Vec{\xi} (\coo) \equiv {\bf u}(\coo)
\sqrt{m/[2k_{\rm B}\tilde{T}(\coo)]}$ 
where 
$m{\bf u}(\coo) \equiv {\bf p} - m\overline{\bf v}(\coo)$, 
and to introduce the dimensionless collision operator 
\begin{align}
\hat{L}_{\sigma}^{'} \left[ \Psi_{\sigma}(1) \right ] \equiv 
&  
\int d\Vec{\xi}_{2}
\int d\Vec{\xi}_{3}
\int d\Vec{\xi}_{4}
\delta \left ( 
\Vec{\xi}_{1} + \Vec{\xi}_{2} - \Vec{\xi}_{3} - \Vec{\xi}_{4} \right )
\delta 
\left ( \xi_{1}^{2} + \xi_{2}^{2} - \xi_{3}^{2} - \xi_{4}^{2} \right )
\nonumber
\\
& 
\times 
\left [ 1- \tilde{f}_{\sigma}(1) \right]
\left [ 1- \tilde{f}_{- \sigma}(2) \right] 
\tilde{f}_{- \sigma}(3)
\tilde{f}_{\sigma}(4) 
\left [ 
\Psi_{\sigma}(4) + \Psi_{-\sigma}(3)
- \Psi_{- \sigma}(2) - \Psi_{\sigma}(1)
\right ]. 
\label{def-Ldash}
\end{align}
The collision integral $\hat{L}_{\sigma} [\Psi_{\sigma}(1)]$ is then reduced to 
$\hat{L}_{\sigma} [\Psi_{\sigma}(1)] = 
\hat{L}_{\sigma}^{'} \left[ \Psi_{\sigma}(1) \right ]/\tilde{C}(\coo)$, 
where the coefficient $\tilde{C}(\coo)$ is defined as 
$
\tilde{C}(\coo) \equiv 
\hbar^{3}\pi^{3}/ \left \{ 4a^{2}m \left [ k_{\rm B}\tilde{T}(\coo) \right]^{2} \right \}
$.

Here, let us substitute the distribution function in the local equilibrium 
to the left hand side of the Boltzmann equation: 
\begin{align}
&
\left[ 
\frac{\partial }{\partial t} 
+ 
\frac{\bf p}{m}\cdot \nabla_{\bf r}
- 
\nabla U_{\sigma}(\coo)\cdot \nabla_{\bf p}
\right]
\tilde{f}_{\sigma}({\bf p}, \coo)
\nonumber
\\
=& 
\left [ 
\frac{1}{z_{\sigma}(\coo)}
\left ( 
\frac{\partial}{\partial t}
+ \frac{\bf p}{m}\cdot\nabla_{\bf r}
\right )
z_{\sigma}(\coo) 
+ 
\frac{mu^{2}(\coo)}{2k_{\rm B}\tilde{T}^{2}(\coo)}
\left ( 
\frac{\partial}{\partial t}
+ \frac{\bf p}{m}\cdot\nabla_{\bf r}
\right ) 
\tilde{T}(\coo) 
\right .
\nonumber
\\
& 
\left .
+ 
\frac{m{\bf u}(\coo)}{k_{\rm B}\tilde{T}(\coo)}
\cdot 
\left ( 
\frac{\partial}{\partial t}
+ \frac{\bf p}{m}\cdot\nabla_{\bf r}
\right ) 
\overline{\bf v}(\coo)
+ 
\tilde{\beta}(\coo) 
\nabla_{\bf r}U_{\sigma}(\coo) \cdot {\bf u}(\coo)
\right ]
\nonumber
\\
&
\qquad\qquad 
\times
\left [ 
1 - \tilde{f}_{\sigma}({\bf p}, \coo) 
\right ]
\tilde{f}_{\sigma}({\bf p}, \coo). 
\label{kinetic-term1}
\end{align} 
This equation can be written in a simpler form as shown below. 
Note that density and pressure satisfy the following equations: 
\begin{align} 
\frac{\partial \tilde{n}_{\sigma}(\coo)}{\partial t} 
=& 
\frac{3}{2} 
\frac{\tilde{n}_{\sigma}(\coo)}{\tilde{T}(\coo)}
\frac{\partial \tilde{T}(\coo)}{\partial t} 
+ 
\frac{\gamma_{\sigma}(\coo)k_{\rm B}\tilde{T}(\coo)}{z_{\sigma}(\coo)}
\frac{\partial z_{\sigma}(\coo)}{\partial t}
\label{densDifferentialT1}
\\
=& 
- 
\frac{3}{2} 
\frac{\tilde{n}_{\sigma}(\coo)}{\tilde{T}(\coo)} 
\overline{\bf v}(\coo) 
\cdot 
\nabla_{\bf r} \tilde{T}(\coo)
-  
\frac{\gamma_{\sigma}(\coo)k_{\rm B}\tilde{T}(\coo)}{z_{\sigma}(\coo)} 
\overline{\bf v}(\coo) 
\cdot 
\nabla_{\bf r}z_{\sigma}(\coo) 
\nonumber
\\
&
\qquad
- 
\tilde{n}_{\sigma}(\coo) 
\left [ 
\nabla_{\bf r} \cdot 
{\bf v}(\coo) 
\right ],  
\label{densDifferentialT2}
\\
\sum\limits_{\sigma}
\frac{\partial \tilde{P}_{\sigma}(\coo)}{\partial t} 
=& 
\sum\limits_{\sigma} 
\left \{
- 
\frac{5}{2} \frac{\tilde{P}_{\sigma}(\coo)}{\tilde{T}(\coo)} 
\nabla_{\bf r} \tilde{T}(\coo)
\cdot 
\overline{\bf v}(\coo)
-  
\frac{\tilde{n}_{\sigma}(\coo) k_{\rm B} \tilde{T}(\coo)}{z_{\sigma}(\coo)}
\nabla_{\bf r} z_{\sigma}(\coo)
\cdot 
\overline{\bf v}(\coo)
\right. 
\nonumber
\\
&
\left .
\qquad 
- 
\frac{5}{3} 
\tilde{P}_{\sigma}(\coo)
\left [
\nabla_{\bf r} \cdot
 \overline{\bf v}(\coo)
\right ] 
\right \}, 
\label{PressDifferentialT1}
\\
\frac{\partial \tilde{P}_{\sigma}(\coo)}{\partial t}
&= 
\frac{5}{2} \frac{\tilde{P}_{\sigma}(\coo)}{\tilde{T}(\coo)}
\frac{\partial \tilde{T}(\coo)}{\partial t} 
+ 
\frac{\tilde{n}_{\sigma}(\coo) k_{\rm B} \tilde{T}(\coo)}{z_{\sigma}(\coo)}
\frac{\partial z_{\sigma}(\coo)}{\partial t}, 
\label{PressDifferentialT2}
\\
\nabla_{\bf r}\tilde{P}_{\sigma}(\coo)
&= 
\frac{5}{2} \frac{\tilde{P}_{\sigma}(\coo)}{\tilde{T}(\coo)} 
\nabla_{\bf r} \tilde{T}(\coo)
+ 
\frac{\tilde{n}_{\sigma}(\coo) k_{\rm B} \tilde{T}(\coo)}{z_{\sigma}(\coo)}
\nabla_{\bf r} z_{\sigma}(\coo), 
\label{PressDifferentialX}
\end{align}
where we define 
$
\gamma_{\sigma}(\coo) \equiv 
{\mathcal F}_{1/2}(z_{\sigma}(\coo))
/
[k_{\rm B}\tilde{T}(\coo)\Lambda^{3}(\coo)] 
$. 
Here, we shall consider the following equation: 
\begin{align}
\sum\limits_{\sigma} 
\left [
\frac{\tilde{n}_{\sigma}(\coo)}{\gamma_{\sigma}(\coo)}
\frac{\partial \tilde{n}_{\sigma}(\coo)}{\partial t}
- 
\frac{\partial \tilde{P}_{\sigma}(\coo)}{\partial t}
\right ]. 
\label{derivable-eqT}
\end{align}
Using Eqs. (\ref{densDifferentialT1}) - (\ref{PressDifferentialT2}), 
and (\ref{derivable-eqT}), 
one obtains 
\begin{align}
\frac{\partial \tilde{T}(\coo)}{\partial t} 
= 
- 
\overline{\bf v}(\coo) \cdot \nabla_{\bf r} \tilde{T}(\coo) 
- \frac{2}{3}
\tilde{T}(\coo) 
\left [ 
\nabla_{\bf r} \cdot 
\overline{\bf v}(\coo) 
\right ].  
\label{TempDifferentialT}
\end{align}
From Eqs. (\ref{densDifferentialT1}), (\ref{PressDifferentialX}) 
and (\ref{TempDifferentialT}), 
one also obtains the following equation: 
\begin{align}
\frac{\partial z_{\sigma}(\coo)}{\partial t}
=
- 
\overline{\bf v}(\coo) 
\cdot 
\nabla_{\bf r}z_{\sigma}(\coo) 
= 
\frac{z_{\sigma}(\coo)}{\tilde{n}_{\sigma}(\coo) k_{\rm B} \tilde{T}(\coo)} 
\overline{\bf v}(\coo) 
\cdot 
\left [ 
\nabla_{\bf r}\tilde{P}_{\sigma}(\coo)
- 
\frac{5}{2} \frac{\tilde{P}_{\sigma}(\coo)}{\tilde{T}(\coo)} 
\nabla_{\bf r} \tilde{T}(\coo)
\right ]. 
\label{fugaDifferentialT}
\end{align} 
Using the above equations, 
we reduce the left hand side of Boltzmann equation to 
\begin{align}
&
\left[ 
\frac{\partial }{\partial t} 
+ 
\frac{\bf p}{m}\cdot \nabla_{\bf r}
- 
\nabla U_{\sigma}(\coo)\cdot \nabla_{\bf p}
\right]
\tilde{f}_{\sigma}({\bf p}, \coo)
\nonumber
\\
=& 
\left (
\frac{1}{\tilde{T}(\coo)}
{\bf u}(\coo) \cdot \nabla_{\bf r}\tilde{T}(\coo)
\left [
\frac{mu^{2}(\coo)}{2k_{\rm B}\tilde{T}(\coo)} 
- 
\frac{5}{2} 
\frac{{\mathcal F}_{5/2}(z_{\sigma}(\coo)) }{{\mathcal F}_{3/2}(z_{\sigma}(\coo))}
\right ]
\right . 
\nonumber
\\
& 
\left .
+ 
\frac{m}{k_{\rm B}\tilde{T}(\coo)} 
\sum\limits_{\mu\nu}
D_{\mu\nu}(\coo)
\left [
u_{\mu}(\coo)u_{\nu}(\coo)
- 
\delta_{\mu\nu}\frac{1}{3}u^{2}(\coo)
\right ]
+
\frac{\tilde{n}_{\rm tot}}{\tilde{n}_{\sigma}}
{\bf d}_{\sigma}(\coo)\cdot {\bf u}(\coo)
\right ) 
\nonumber
\\
& \qquad \qquad \times 
\left [ 
1 - \tilde{f}_{\sigma}({\bf p}, \coo) 
\right ]
\tilde{f}_{\sigma}({\bf p}, \coo), 
\label{kinetic-term2}
\end{align}
where we define ${\bf d}_{\sigma}(\coo)$ as 
\begin{align}
{\bf d}_{\sigma}(\coo)
&\equiv 
\frac{1}{k_{\rm B}\tilde{T}(\coo) \tilde{n}_{\rm tot}}
\frac{\tilde{n}_{\sigma}\tilde{n}_{-\sigma}}{\tilde{n}_{\rm tot}}
\left \{
\left [
\frac{
\nabla \tilde{P}_{\sigma}(\coo)
}
{
\tilde{n}_{\sigma}
}
+ 
\nabla U_{\sigma}(\coo)
\right ]
- 
\left [
\frac{
\nabla \tilde{P}_{-\sigma}(\coo)
}
{
\tilde{n}_{-\sigma}
}
+ 
\nabla U_{-\sigma}(\coo)
\right ]
\right \}. 
\label{differntial-Force}
\end{align}
In the population balanced gas, one finds ${\bf d}_{\sigma}(\coo)=0$. 
As a result, the left hand side of the Boltzmann equation under the local equilibrium 
in the population balanced Fermi gas is reduced to 
\begin{align}
&
\left[ 
\frac{\partial }{\partial t} 
+ 
\frac{\bf p}{m}\cdot \nabla_{\bf r}
- 
\nabla U_{\sigma}(\coo)\cdot \nabla_{\bf p}
\right]
\tilde{f}_{\sigma}({\bf p}, \coo)
\nonumber
\\
=& 
\left \{
\sqrt{\frac{2k_{\rm B}\tilde{T}(\coo)}{m}}
\frac{\Vec{\xi}(\coo) \cdot \nabla_{\bf r}\tilde{T}(\coo)}{\tilde{T}(\coo)}
\left [
\xi^{2}
- 
\frac{5}{2} 
\frac{{\mathcal F}_{5/2}(z_{\sigma}(\coo)) }{{\mathcal F}_{3/2}(z_{\sigma}(\coo))}
\right ]
\right . 
\nonumber
\\
& 
\left .
+ 
2
\sum\limits_{\mu\nu}
D_{\mu\nu}(\coo)
\left [
\xi_{\mu}(\coo)\xi_{\nu}(\coo)
- 
\delta_{\mu\nu}\frac{1}{3}\xi^{2}(\coo)
\right ]
\right \}
\left [ 
1 - \tilde{f}_{\sigma}({\bf p}, \coo) 
\right ]
\tilde{f}_{\sigma}({\bf p}, \coo). 
\label{Enskog1FinalEq}
\end{align}

We introduce an ansatz for the departure from the equilibrium, which is 
\begin{align}
\Psi_{\sigma}(\xi) 
\equiv 
&
\tilde{C}(\coo)
\left \{ 
\left [
\frac{2k_{\rm B}\tilde{T}(\coo)}{m}
\right ]^{1/2}
\frac{\nabla\tilde{T}(\coo)\cdot \Vec{\xi}(\coo)}{\tilde{T}(\coo)} 
A_{\sigma}(\xi) 
\nonumber
\right. 
\\ 
&
\left.
+ 
2 \sum\limits_{\mu\nu}D_{\mu\nu}(\coo)
\left [ 
\xi_{\mu}(\coo)\xi_{\nu}(\coo) 
- 
\frac{1}{3}\delta_{\mu\nu} \xi^{2}(\coo)
\right ]
B_{\sigma}(\xi) 
\right \}. 
\label{ansatz}
\end{align} 
This comes from a consideration that the solution $\Psi_{\sigma}(\xi)$ must be a linear function of 
$\nabla \tilde{T}({\bf r}, t)$ and $D_{\mu\nu} ({\bf r}, t)$, based on Eq. (\ref{Enskog1FinalEq}). 
We substitute this into the collision integral on the right hand side of the Boltzmann equation. 
Comparing Eq. (\ref{Enskog1FinalEq}) with this result, 
one obtains the following relations: 
\begin{align}
&
\hat{L}_{\sigma}^{'}
\left [
\Vec{\xi}
A_{\sigma}(\Vec{ \xi})
\right ] 
= 
\Vec{\xi}
\left [
\xi^{2}
- 
\frac{5}{2} 
\frac{{\mathcal F}_{5/2}(z_{\sigma}(\coo)) }{{\mathcal F}_{3/2}(z_{\sigma}(\coo))}
\right ]
\left [ 
1 - \tilde{f}_{\sigma}({\bf p}, \coo) 
\right ]
\tilde{f}_{\sigma}({\bf p}, \coo),  
\label{Ldash-A}
\\
&
\hat{L}_{\sigma}^{'}
\left [
\left [ 
\xi_{\mu}(\coo)\xi_{\nu}(\coo) 
- 
\frac{1}{3}\delta_{\mu\nu} \xi^{2}(\coo)
\right ]
B_{\sigma}(\xi) 
\right ]
\nonumber
\\ 
& \qquad \qquad 
=
\left [ 
\xi_{\mu}(\coo)\xi_{\nu}(\coo) 
- 
\frac{1}{3}\delta_{\mu\nu} \xi^{2}(\coo)
\right ]
\left [ 
1 - \tilde{f}_{\sigma}({\bf p}, \coo) 
\right ]
\tilde{f}_{\sigma}({\bf p}, \coo).  
\label{Ldash-B}
\end{align}

THe ansatz (\ref{ansatz}) automatically satisfies two constraints (\ref{constraint1}) and (\ref{constraint3}). 
For the constraint (\ref{constraint2}) to be satisfied, 
the function $A_{\sigma}(\xi)$ should satisfy 
\begin{align}
\sum\limits_{\sigma} 
\intxi (\coo)
\tilde{f}_{\sigma} (\Vec{\xi}, \coo)
\left [
1- \tilde{f}_{\sigma} (\Vec{\xi}, \coo)
\right ] 
\xi^{2}(\coo)
A_{\sigma}(\xi) 
= 
0. 
\label{constraintA}
\end{align}

Transport coefficients such as the thermal conductivity and the viscosity are obtained 
using the ansatz (\ref{ansatz}). 
The thermal conductivity $\kappa$ is defined by 
\begin{align}
{\bf Q}(\coo) 
&= 
-\kappa (\coo) \nabla \tilde{T}(\coo). 
\label{heatcurrentDef}
\end{align}
From Eqs. (\ref{heatcurrent}), (\ref{discrepancy-from-localeq}), and (\ref{ansatz}), it is given by 
\begin{align}
\kappa (\coo) 
&= 
- 
\frac{k_{\rm B}}{24a^{2}}
\left [ 
\frac{2k_{\rm B}\tilde{T}(\coo)}{m}
\right ]^{1/2} 
\sum\limits_{\sigma}
\int d\Vec{\xi}
\xi^{4}(\coo) 
\left [ 
1- \tilde{f}_{\sigma}({\bf p}, \coo)
\right ]
\tilde{f}_{\sigma}({\bf p}, \coo) 
A_{\sigma}(\xi). 
\label{heatcurrentExp}
\end{align}
The shear viscosity $\eta$ is defined by 
\begin{align}
P(\coo) 
= 
\delta_{\mu\nu}\tilde{P}_{\mu\nu}(\coo)
-
2 
\eta(\coo) 
\left [ 
D_{\mu\nu}(\coo) 
-\frac{1}{3}
{\rm Tr} D(\coo)\delta_{\mu\nu}
\right ]. 
\label{viscosityDef}
\end{align}
From Eqs. (\ref{pressure}), (\ref{discrepancy-from-localeq}), and (\ref{ansatz}), 
it is given by 
\begin{align}
\eta(\coo) 
\equiv
-\frac{m}{60a^{2}}
\left [
\frac{2k_{\rm B}\tilde{T}(\coo)}{m}
\right ]^{1/2}
\sum\limits_{\sigma}
\intxi 
\xi^{4}(\coo)
\left [ 
1 - \tilde{f}_{\sigma}({\bf p},\coo) 
\right ]
\tilde{f}_{\sigma}({\bf p},\coo) 
B_{\sigma}(\xi). 
\label{viscosityExp}
\end{align}
%%%%%%%%%%
Note that the second viscosity (the bulk viscosity) is absent. 
In more general, 
the second viscosity vanishes in the normal gas interacting with the $s$-wave scattering, 
because a uniform compression at a steady rate changes the thermodynamic equilibrium 
into a new one (see the second of Ref.~\cite{FermiLiquidBOOK}).  

Using these quantities, 
we reduce hydrodynamic equations for the velocity and the energy density to 
\begin{align}
&
mn_{\rm tot}(\coo)
\left [
\frac{\partial }{\partial t}
+ 
{\bf v}(\coo)\cdot \nabla
\right ]
v_{\mu}
+ 
\frac{\partial}{\partial x_{\mu}}\tilde{P}(\coo)
+
\frac{\partial}{\partial x_{\mu}}
\left [
gn_{\uparrow}(\coo)n_{\downarrow}(\coo)
\right ]
\nonumber 
\\
&
\qquad \qquad 
=  
\frac{\partial}{\partial x_{\nu}}
\left \{ 
2 
\eta(\coo)
\left [ 
D_{\mu\nu}(\coo) 
-\frac{1}{3}
{\rm Tr} D(\coo)\delta_{\mu\nu}
\right ] 
\right \}, 
\label{velocityEq-with-viscosity}
\\
&
\frac{\partial }{\partial t}E(\coo)
+ 
\nabla 
\left [ 
E(\coo){\bf v}(\coo)
\right ]
+ 
\left [
\nabla\cdot {\bf v}(\coo)
\right ]
\tilde{P}(\coo)
\nonumber 
\\
&
\qquad \qquad 
=
\nabla 
\left [ 
\kappa (\coo)\nabla T(\coo)
\right ]
+ 
2 
\eta(\coo)
\sum\limits_{\mu\nu}
\left [ 
D_{\mu\nu}(\coo) 
-\frac{1}{3}
{\rm Tr} D(\coo)\delta_{\mu\nu}
\right ]^{2}.  
\label{energydensityEq-with-heatcurent}
\end{align}

Based on Refs.~\cite{Uehling1933,Nikuni1998}, 
we shall take the function $A_{\sigma}(\xi)$ as 
\begin{align}
A_{\sigma}(\xi) 
= 
A 
\left [ 
\xi^{2}(\coo) -\frac{5}{2} 
\frac{{\mathcal F}_{5/2}(z_{\sigma})}{{\mathcal F}_{3/2}(z_{\sigma})}
\right ]. 
\end{align} 
Note that this satisfies a constraint in Eq. (\ref{constraintA}). 
Multiplying 
Eq. (\ref{Ldash-A}) by 
$\Vec{\xi} [ \xi^{2}-\frac{5}{2}
{\mathcal F}_{5/2}(z_{\sigma})/{\mathcal F}_{3/2}(z_{\sigma})]$, 
and integrating over $\Vec{\xi}$, 
one obtains the coefficient $A$ as follows: 
\begin{align}
A = 
\frac{
\sum\limits_{\sigma}
\intxi 
\xi^{2}
\left [ \xi^{2} - \frac{5}{2}
\frac{{\mathcal F}_{5/2}(z_{\sigma})}{{\mathcal F}_{3/2}(z_{\sigma})}
\right ]^{2}
[1-\tilde{f}_{\sigma}]\tilde{f}_{\sigma}
}
{
\sum\limits_{\sigma}
\intxi 
\left [ 
\left ( 
\xi^{2} - \frac{5}{2}
\frac{{\mathcal F}_{5/2}(z_{\sigma})}{{\mathcal F}_{3/2}(z_{\sigma})}
\right )
\Vec{\xi}
\right ]
\hat{L}' 
\left [ 
\Vec{\xi}
\left ( \xi^{2} - \frac{5}{2}
\frac{{\mathcal F}_{5/2}(z_{\sigma})}{{\mathcal F}_{3/2}(z_{\sigma})}
 \right )
\right ]
}
= 
\frac{15}{4}
\frac{\pi^{3/2}}{{\mathcal I}_{\rm A}}
\sum\limits_{\sigma} 
\left [ 
\frac{7}{2} 
{\mathcal F}_{7/2}(z_{\sigma})
- 
\frac{5}{2} 
\frac{{\mathcal F}_{5/2}^{2}(z_{\sigma})}{{\mathcal F}_{3/2}(z_{\sigma})}
\right ], 
\label{obtainedA}
\end{align}
where ${\mathcal I}_{\rm A}$ is defined as 
\begin{align}
{\mathcal I}_{\rm A}
\equiv 
\sum\limits_{\sigma}
\intxi 
\left [ 
\left ( 
\xi^{2} - \frac{5}{2}
\frac{{\mathcal F}_{5/2}(z_{\sigma})}{{\mathcal F}_{3/2}(z_{\sigma})} 
\right )
\Vec{\xi}
\right ]
\hat{L}' 
\left [ 
\Vec{\xi}
\left ( \xi^{2} - \frac{5}{2} 
\frac{{\mathcal F}_{5/2}(z_{\sigma})}{{\mathcal F}_{3/2}(z_{\sigma})}
\right )
\right ]. 
\end{align}

On the other hand, 
based on Refs.~\cite{Uehling1933,Nikuni1998}, 
we shall take the function $B_{\sigma}(\xi)$ as $B_{\sigma}(\xi) \equiv B$. 
Integrating over $\Vec{\xi}$ and summing over $\sigma$, $\nu$ and $\mu$, 
after multiplying Eq. (\ref{Ldash-B}) by $(\xi_{\nu}\xi_{\mu}-\frac{1}{3}\delta_{\nu\mu}\xi^{2})$, 
one obtains the coefficient $B$ as follows: 
\begin{align}
B 
= & 
\frac{
\sum\limits_{\sigma}
\sum\limits_{\mu\nu}
\intxi 
\left [
\xi_{\mu}(\coo)\xi_{\nu}(\coo)
-\frac{1}{3}\delta_{\mu\nu}\xi^{2}(\coo)
\right ]^{2}
\tilde{f}_{\sigma}({\bf p}, \coo)
\left [
1-\tilde{f}_{\sigma}({\bf p}, \coo) 
\right ]
}
{
\sum\limits_{\sigma}
\sum\limits_{\mu\nu}
\intxi 
\left [
\xi_{\mu}(\coo)\xi_{\nu}(\coo)
-\frac{1}{3}\delta_{\mu\nu}\xi^{2}(\coo)
\right ]
\hat{L}_{\sigma}' 
\left [
\left (
\xi_{\mu}(\coo)\xi_{\nu}(\coo)
-\frac{1}{3}\delta_{\mu\nu}\xi^{2}(\coo)
\right )
\right ]
}
\nonumber 
\\
=  &
\frac{5}{2} 
\pi^{3/2}
\frac{1}
{{\mathcal I}_{\rm B}}
\sum\limits_{\sigma} 
{\mathcal F}_{5/2} (z_{\sigma}), 
\label{obtainedB}
\end{align}
where 
\begin{align}
{\mathcal I}_{\rm B} 
\equiv
\sum\limits_{\sigma}\intxi 
\left [
\xi_{\mu}(\coo)\xi_{\nu}(\coo)
-\frac{1}{3}\delta_{\mu\nu}\xi^{2}(\coo)
\right ]
\hat{L}_{\sigma}' 
\left [
\left (
\xi_{\mu}(\coo)\xi_{\nu}(\coo)
-\frac{1}{3}\delta_{\mu\nu}\xi^{2}(\coo)
\right )
\right ]. 
\end{align}

The collision integral satisfies the hermitian property 
\begin{align}
\sum\limits_{\sigma}
\int d\Vec{\xi}
\Phi_{\sigma}(\Vec{\xi})
\hat{L}'_{\sigma}[\Psi_{\sigma}(\Vec{\xi})]
=  
\sum\limits_{\sigma}
\int d\Vec{\xi}
\Psi_{\sigma}(\Vec{\xi})
\hat{L}'_{\sigma}[\Phi_{\sigma}(\Vec{\xi})]. 
\end{align}
The collision integral in Eq. (\ref{def-Ldash}) also satisfies 
$\hat{L}_{\sigma}'[\Vec{\xi}_{1}] = 0$, and $\hat{L}_{\sigma}'[\xi_{1}^{2}] = 0$ 
owing to the conservation of the momentum and the energy; 
then, ${\mathcal I}_{\rm A}$ and ${\mathcal I}_{\rm B}$ are reduced to simpler formulae given by 
\begin{align}
{\mathcal I}_{\rm A}
= 
\sum\limits_{\sigma}
\intxi 
\Vec{\xi}\xi^{2}
\hat{L}'_{\sigma}
\left [
\xi^{2}\Vec{\xi}
\right ] 
, \qquad &
{\mathcal I}_{\rm B}
= 
\sum\limits_{\sigma}
\intxi 
\xi_{\mu}\xi_{\nu}
\hat{L}'_{\sigma}
\left [
\xi_{\mu}\xi_{\nu}
\right ]. 
\label{Eq58}
\end{align} 
We shall introduce new variables given by 
$\V{\xi}_{1} \equiv (\V{\xi}_{0}  + \V{\xi}')/\sqrt{2}$, 
$\V{\xi}_{2} \equiv (\V{\xi}_{0}  - \V{\xi}')/\sqrt{2}$, 
$\V{\xi}_{3} \equiv (\V{\xi}_{0}'  + \V{\xi}'')/\sqrt{2}$, 
and 
$\V{\xi}_{4} \equiv (\V{\xi}_{0}'  - \V{\xi}'')/\sqrt{2}$. 
Note that these variables satisfy relations $\V{\xi}_{0} = \V{\xi}_{0}'$ and $|\V{\xi}'| = |\V{\xi}''|$, 
because of the conservation of the momentum of the center of mass and of the energy in the relative motion. 
Here, we shall define a function $F_{\sigma}(\xi_{0},\xi',y',y'')$ given by 
\begin{align}
F_{\sigma}(\xi_{0},\xi',y',y'')
&\equiv 
\tilde{f}_{\sigma}(1) \tilde{f}_{-\sigma}(2) 
\left [ 1- \tilde{f}_{-\sigma}(3) \right ]
\left [ 1- \tilde{f}_{\sigma}(4) \right ]
\\
&= 
\frac{z_{\sigma}z_{-\sigma} \exp{[-(\xi_{0}^{2}+\xi'^{2})]}}
{[1+z_{\sigma}\exp{(-\xi_{1}^{2})}]
[1+z_{-\sigma}\exp{(-\xi_{2}^{2})}]
[1+z_{-\sigma}\exp{(-\xi_{3}^{2})}]
[1+z_{\sigma}\exp{(-\xi_{4}^{2})}]}, 
\end{align}
where 
$\xi_{1}^{2} \equiv (\xi_{0}^{2} + 2 \xi_{0}\xi'y' +\xi'^{2})/2$, 
$\xi_{2}^{2}  \equiv (\xi_{0}^{2} - 2 \xi_{0}\xi'y' +\xi'^{2})/2$, 
$\xi_{3}^{2}  \equiv (\xi_{0}^{2} + 2 \xi_{0}\xi'y'' +\xi'^{2})/2$, 
and
$\xi_{4}^{2}  \equiv (\xi_{0}^{2} - 2 \xi_{0}\xi'y'' +\xi'^{2})/2$. 
The function $F_{\sigma}(\xi_{0}, \xi', y', y'')$ satisfies the following relations: 
\begin{align}
\sum\limits_{\sigma}
F_{\sigma}(\xi_{0},\xi', y'',y') 
= &
\sum\limits_{\sigma}
F_{-\sigma}(\xi_{0},\xi', y',y'') 
\\
\sum\limits_{\sigma}F_{\sigma}(\xi_{0},\xi', -y',-y'') 
=& 
\sum\limits_{\sigma}F_{\sigma}(\xi_{0},\xi', y',y'') 
\\
\sum\limits_{\sigma}F_{\sigma}(\xi_{0},\xi', -y',y'') 
=& 
\sum\limits_{\sigma}F_{\sigma}(\xi_{0},\xi', y',-y''). 
\end{align}
The integrals in (\ref{Eq58}) in terms of the new variables are thus rewritten as 
\begin{align}
{\mathcal I}_{A}
=& 
\sqrt{2}\pi^{3} 
\int_{0}^{\infty}d\xi_{0} \xi_{0}^{4} 
\int_{0}^{\infty}d\xi' \xi'^{7} 
\int_{-1}^{1}dy'\int_{-1}^{1}dy'' 
\sum\limits_{\sigma}F_{\sigma}(\xi_{0},\xi';y',y'')
(y'^{2}+y''^{2}-2y'^{2}y''^{2}), 
\\
{\mathcal I}_{B}
=& 
\frac{1}{\sqrt{2}}
\pi^{3}
\int_{0}^{\infty}d\xi_{0} \xi_{0}^{2}
\int_{0}^{\infty}d\xi' \xi'^{7}
\int_{-1}^{1}dy'
\int_{-1}^{1}dy'' 
\sum\limits_{\sigma}F_{\sigma}(\xi_{0},\xi';y',y'') 
(1+y'^{2}+y''^{2}-3 y'^{2}y''^{2}). 
\end{align}

For simplicity, 
we again introduce variables $\xi_{0}  \equiv \sqrt{2\eta} \cos{\phi}$ 
and 
$\xi'  \equiv \sqrt{2\eta} \sin{\phi}$. 
The condition of the population balanced gas is given by $\tilde{\mu}_{\uparrow} = \tilde{\mu}_{\downarrow}$. 
In this condition, 
the function $F_{\sigma}(\eta,\phi,y',y'')$ can be reduced to 
\begin{align}
F_{\sigma}(\eta,\phi,y',y'')
= &
\frac{1}{4}
\,\,
\frac{1}
{
\cosh{(\eta-\tilde{\beta} \tilde{\mu}_{\rm tot}/2)}
+ 
\cosh{(\eta y'\sin{2\phi})}
}
\,\,
\frac{1}
{
\cosh{(\eta-\tilde{\beta}\tilde{\mu}_{\rm tot}/2)}
+ 
\cosh{(\eta y''\sin{2\phi})}
}, 
\end{align}
where $\tilde{\mu}_{\rm tot} \equiv \tilde{\mu}_{\uparrow} + \tilde{\mu}_{\downarrow}$. 
As a result, 
integrals ${\mathcal I}_{A}$ and ${\mathcal I}_{B}$ are reduced to 
\begin{align}
{\mathcal I}_{A}
=& 
2^{5}\pi^{3}
\int_{0}^{\infty}d\eta \eta^{11/2}
\int_{0}^{\pi/2}d\phi
\cos^{4}{\phi}\sin^{7}{\phi}
\int_{-1}^{1}dy' 
\int_{-1}^{1}dy'' 
(y'^{2}+y''^{2}-2y'^{2}y''^{2})
\nonumber
\\
&\times
\frac{1}
{
\cosh{(\eta-\tilde{\beta} \tilde{\mu}_{\rm tot}/2)}
+ 
\cosh{(\eta y'\sin{2\phi} )}
}
\,\,
\frac{1}
{
\cosh{(\eta-\tilde{\beta}\tilde{\mu}_{\rm tot}/2)}
+ 
\cosh{(\eta y''\sin{2\phi}  )}
}, 
\label{EquationIA}
\\
{\mathcal I}_{B}
=& 
2^{3}\pi^{3}
\int_{0}^{\infty}d\eta \eta^{9/2}
\int_{0}^{\pi/2}d\phi
\cos^{2}{\phi}\sin^{7}{\phi}
\int_{-1}^{1}dy' 
\int_{-1}^{1}dy'' 
(1+y'^{2}+y''^{2}-3y'^{2}y''^{2})
\nonumber
\\
&\times
\frac{1}
{
\cosh{(\eta-\tilde{\beta}\tilde{\mu}_{\rm tot}/2)}
+ 
\cosh{(\eta y'\sin{2\phi} )}
}
\,\,
\frac{1}
{
\cosh{(\eta-\tilde{\beta}\tilde{\mu}_{\rm tot}/2)}
+ 
\cosh{(\eta y''\sin{2\phi}  )}
}. 
\label{EquationIB}
\end{align}

Before closing this section, 
we summarize that the thermal conductivity $\kappa$ and the viscosity $\eta$ are 
given by 
\begin{align}
\kappa
=& 
-\frac{k_{\rm B}}{a^{2}}
\left ( 
\frac{2k_{\rm B}\tilde{T}(\coo)}{m}
\right )^{1/2}
\frac{75}{128}
\pi^{3}
\frac{1}{{\mathcal I}_{\rm A}}
\left \{
\sum\limits_{\sigma}
\left [
\frac{7}{2}
{\mathcal F}_{7/2}(z_{\sigma})
- 
\frac{5}{2}
\frac{{\mathcal F}_{5/2}^{2}(z_{\sigma})}{{\mathcal F}_{3/2}(z_{\sigma})}
\right ]
\right \}^{2}, 
\label{EC-HeatCurrent}
\\
\eta
=& - 
\frac{m}{a^{2}}
 \left ( 
\frac{2k_{\rm B}\tilde{T}(\coo)}{m}
\right )^{1/2}
\frac{5}{32}\pi^{3}
\frac{1}{{\mathcal I}_{\rm B}}
\left [
\sum\limits_{\sigma}
{\mathcal F}_{5/2}(z_{\sigma})
\right ]^{2}, 
\label{EC-Viscous}
\end{align}
where Eqs. (\ref{heatcurrentExp}), (\ref{viscosityExp}), (\ref{obtainedA}) and (\ref{obtainedB}) are used.

\section{Relaxation Time}\label{relaxation}

A purpose of the present appendix is to derive relaxation times using results obtained in the previous section. 
Let us consider the solution in the collisional hydrodynamic regime. 
In this regime, the departure from local equilibrium 
on the left hand side of the linearized Boltzmann equation (\ref{linearized+deltamu}) are neglected. 
Solving it for $\delta \nu_{\sigma}({\bf p}, {\bf q}, \omega)$, 
one obtains 
\begin{eqnarray}
\delta \nu_{\sigma} ({\bf p}, {\bf q}, \omega)
&=&
i \tau 
\left \{
\left ( 
\omega -\frac{{\bf p}\cdot{\bf q}}{m}
\right )
\left [
a_{\sigma}({\bf q}, \omega)
+ 
{\bf b}({\bf q}, \omega) \cdot {\bf p}
+ 
c({\bf q}, \omega)p^{2}
\right ]
\right. 
\nonumber
\\
&&
\left.
+ 
\frac{{\bf p}\cdot{\bf q}}{m}
g 
\left [ 
a_{-\sigma}({\bf q}, \omega)
W_{-\sigma, 0}
+
c({\bf q}, \omega)W_{-\sigma, 2}
\right ]
\right \}. 
\label{delta-nu-ABC}
\end{eqnarray}

A closed set of equations for $a_{\sigma}({\bf q}, \omega)$, 
${\bf b}({\bf q}, \omega)$ and $c({\bf q}, \omega)$ can be obtained 
from Eq. (\ref{linearized+deltamu}) by multiplying Eq. (\ref{linearized+deltamu}) by $1$, ${\bf p}$, $p^{2}$ 
and integrating over ${\bf p}$. 
The zeroth moment yields 
\begin{eqnarray}
\omega a_{\sigma} ({\bf q}, \omega) 
&=&
\frac{W_{\sigma, 2}}{W_{\sigma, 0}}
\left [ 
\frac{{\bf b}({\bf q}, \omega)\cdot{\bf q}}{3m} 
-\omega c({\bf q}, \omega)
\right ]. 
\label{Linearized-Hydro-Desity}
\end{eqnarray}
The second moment yields 
\begin{eqnarray}
\begin{split}
\omega 
a_{\uparrow}({\bf q}, \omega)
W_{\uparrow, 2}
+ 
\omega
a_{\downarrow}({\bf q}, \omega)
W_{\downarrow, 2}
+ 
\left [  
\omega c({\bf q}, \omega) 
- 
\frac{{\bf b}({\bf q}, \omega)\cdot{\bf q}}{3m}
\right ]
\left (
W_{\uparrow, 4} + W_{\downarrow, 4}
\right )
= 0.
\end{split}
\label{Linearized-Hydro-Energy}
\end{eqnarray}
From Eqs. (\ref{Linearized-Hydro-Desity}) and (\ref{Linearized-Hydro-Energy}), 
relations given by 
\begin{eqnarray}
\omega c({\bf q}, \omega)
= 
\frac{{\bf b}({\bf q}, \omega)\cdot{\bf q}}{3m} 
= 
- \frac{\delta{\bf v}({\bf q}, \omega)\cdot{\bf q}}{3m}, 
\label{obtained-linearized-density-energy}
\end{eqnarray} 
and $a_{\sigma} ({\bf q}, \omega) = 0$ are obtained; 
therefore 
the departure from local equilibrium 
in Eq. (\ref{delta-nu-ABC}) 
is reduced into 
\begin{align}
\delta \nu_{\sigma} ({\bf q}, \omega) 
=
i\tau
\frac{{\bf p}\cdot{\bf q}}{2m^{2}}
\beta_{0}
\left [
p^{2}
- 
\frac{
W_{\sigma,4}
}
{W_{\sigma,2}}
\right ]
\delta \theta ({\bf q}, \omega)
+ 
i\tau
\left [ 
-\frac{\delta{\bf v}({\bf q}, \omega)\cdot{\bf q}}{3m}p^{2}
+ 
\frac{{\bf p}\cdot{\bf q}}{m}
{\delta {\bf v}}({\bf q}, \omega)\cdot {\bf p}
\right ]. 
\label{delta-nu-final}
\end{align}

From Eq. (\ref{heatcurrentDef}), 
the heat current in the Fourier representation is given by 
$
{\bf Q}({\bf q}, \omega) 
= -\kappa i {\bf q} \delta T({\bf q}, \omega)
$. 
From Eqs. (\ref{heatcurrent}) and (\ref{delta-nu-final}), 
the thermal conductivity $\kappa$ is thus obtained as 
\begin{align}
\kappa 
= 
- 
\sum\limits_{\sigma} 
k_{\rm B} 
\frac{\beta_{0} \tau}{12m^{4}}
\left (
W_{\sigma,6} -
\frac{
W_{\sigma,4}^{2}
}
{W_{\sigma,2}}
\right ). 
\label{RA-HeatCurrent}
\end{align}

The Fourier representation of the rate-of-strain tensor in Eq. (\ref{rate-of-straintensor}) is given by 
$
D_{\mu\nu} ({\bf q},\omega) 
= i 
\left [ 
q_{\mu} \overline{v}_{\nu} ({\bf q},\omega) 
+ 
q_{\nu} \overline{v}_{\mu} ({\bf q},\omega) 
\right ]/2
$, 
and hence the pressure tensor is 
$
P_{\mu\nu} ({\bf q},\omega) = 
\delta_{\mu,\nu} P({\bf q},\omega) 
-2\eta 
\left [ 
D_{\mu\nu}({\bf q},\omega) 
- 
\delta_{\mu,\nu}
{\rm Tr}D({\bf q},\omega) /3
\right ]
$. 
From Eq. (\ref{delta-nu-final}), 
the viscosity $\eta$ is obtained as 
\begin{align}
\eta = -\sum\limits_{\sigma}\frac{\tau W_{\sigma,4}}{15m^{2}}. 
\label{RA-Viscous}
\end{align}

Comparing Eq. (\ref{EC-HeatCurrent}) with Eq. (\ref{RA-HeatCurrent}), 
we obtain the relaxation time associated with the thermal conductivity 
(denoted by $\tau_{\kappa}$) given by 
\begin{align}
\tau_{\kappa}
=& 
-
15
\frac{\pi^{13/2}\hbar^{7} }{g^{2}m^{3}(k_{\rm B}T)^{2}}
\frac{1}{{\mathcal I}_{\rm A}}
\left \{
\sum\limits_{\sigma} 
\left [ 
\frac{7}{2} 
{\mathcal F}_{7/2}(z_{\sigma})
- 
\frac{5}{2}
\frac{
{\mathcal F}_{5/2}^{2}(z_{\sigma})
}{
{\mathcal F}_{3/2}(z_{\sigma})
}
\right ]
\right \}
= 
-\frac{4\pi^{5}\hbar^{7}}{g^{2}m^{3}(k_{\rm B}T)^{2}}A. 
\label{tau-kappa}
\end{align} 
Comparing Eqs. (\ref{EC-Viscous}) and (\ref{RA-Viscous}), 
on the other hand, 
we obtain the relaxation time associated with viscosity 
(denoted by $\tau_{\eta}$) given by  
\begin{align}
\tau_{\eta}
= &  
-
10 
\frac{\pi^{13/2}\hbar^{7}}{g^{2}m^{3}(k_{\rm B}T)^{2}}
\frac{1}{{\mathcal I}_{\rm B}}
\sum\limits_{\sigma}
{\mathcal F}_{5/2}(z_{\sigma})
= 
-\frac{4\pi^{5}\hbar^{7}}{g^{2}m^{3}(k_{\rm B}T)^{2}}B. 
\label{tau-eta}
\end{align}
We have used the following relation: 
\begin{align}
W_{l,\sigma}
= & 
- (l+1)\frac{1}{\Lambda^{3}}
\frac{(2mk_{\rm B}T)^{l/2}}{k_{\rm B}T} \frac{1}{\sqrt{\pi}} 
\Gamma \left ( \frac{l+1}{2} \right ) 
{\mathcal F}_{\frac{l+1}{2}} (z_{\sigma}). 
\end{align} 
Equations (\ref{EC-HeatCurrent}) and (\ref{EC-Viscous}) 
are written in terms of the temperature and the fugacity in local equilibrium. 
We note that these quantities should be taken as equilibrium values in the expressions for the relaxation times.

On the other hand, 
the mean-collision time $\tau_{\rm coll}$ is defined by 
\begin{align}
\frac{N_{\rm tot}}{V}\frac{1}{\tau_{\rm coll}} \equiv & 
\int \frac{d{\bf p}_{1}}{(2\pi\hbar)^{3}} 
{\mathcal I}_{\rm coll} [f_{\sigma}(1)] 
\\ 
= & 
\frac{2\pi g^{2}}{\hbar}
\int \frac{d{\bf p}_{1}}{(2\pi\hbar)^{3}}
\int \frac{d{\bf p}_{2}}{(2\pi\hbar)^{3}}
\int \frac{d{\bf p}_{3}}{(2\pi\hbar)^{3}}
\int d{\bf p}_{4}
\delta ({\bf p}_{1} + {\bf p}_{2} - {\bf p}_{3} - {\bf p}_{4})
\nonumber
\\
&  
\times 
\delta 
\left (
\frac{p_{1}^{2}}{2m} + \frac{p_{2}^{2}}{2m} 
- \frac{p_{3}^{2}}{2m} - \frac{p_{4}^{2}}{2m}
\right ) 
\left [ 1- f_{\sigma}(1) \right]
\left [ 1- f_{- \sigma}(2) \right] 
f_{- \sigma}(3)
f_{\sigma}(4). 
\end{align} 
Following the procedures analogous to those 
deriving Eqs. (\ref{EquationIA}) and (\ref{EquationIB}), 
we reduce $\tau_{\rm coll}$ to 
\begin{align} 
\frac{N_{\rm tot}}{V} 
\frac{1}{\tau_{\rm coll}} = &
g^{2}
\frac{m^{9/2}(k_{\rm B}T)^{7/2}}{2^{3/2}\pi^{5}\hbar^{10}}
\int_{0}^{\infty}
d\eta \eta^{5/2}
\int_{0}^{\pi/2}d\phi 
\cos^{2}{\phi}
\sin^{3}{\phi} 
\int_{-1}^{1}dy'
\int_{-1}^{1}dy''
\nonumber
\\
&
\times 
\frac{1}
{
\cosh{(\eta-\beta \mu_{\rm tot}/2)}
+ 
\cosh{(\eta y'\sin{2\phi})}
}
\,\,
\frac{1}
{
\cosh{(\eta-\beta \mu_{\rm tot}/2)}
+ 
\cosh{(\eta y''\sin{2\phi})}
}. 
\end{align}

\begin{figure}[htbp]
\begin{center}
\includegraphics[width=7cm,height=7cm,keepaspectratio,clip]{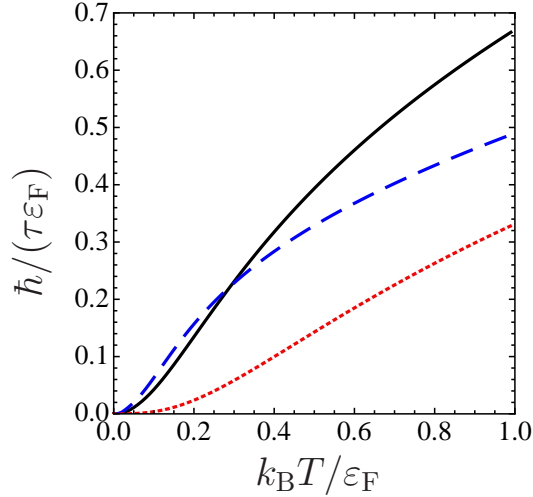}
\end{center}
\caption{
Relaxation rates $1/\tau$ versus temperature. 
Viscous and thermal conductivity relaxation rates are shown with solid and dashed lines, respectively. 
The mean-collision rate is plotted with dotted line. 
A coupling constant $gN_{\rm tot}/V$ is assumed 
to be the Fermi energy: $\alpha \equiv gN_{\rm tot}/V\varepsilon_{\rm F} = 1$.
}
\label{Fig5}
\end{figure}

In Fig.~\ref{Fig5}, 
viscous and thermal relaxation rates are plotted. 
The mean-collision rate is also shown. 
The coupling constant 
$\alpha \equiv gN_{\rm tot}/V \varepsilon_{\rm F} = 1$ is used, 
where $\varepsilon_{\rm F}$ is the Fermi energy. 
Behavior of the viscous relaxation time 
is different from that of the mean-collision time, 
as noted in Ref.~\cite{Vichi2000}. 
The viscous relaxation rate is severalfold bigger than the mean-collision rate, 
and is effective in the hydrodynamic regime compared with other relaxation rates. 
In the low temperature regime, although the thermal relaxation rate is bigger than the viscous one, 
the difference is very small.

In summary, the viscous relaxation rate is the most important in the high temperature regime. 
We apply this viscous relaxation rate to the relaxation time in the moment method, 
because the density oscillation is the most strongly coupled 
with the viscous relaxation, and this relaxation rate is dominant in the high temperature regime.

\section{Random Phase Approximation}\label{RPA}

We solve the linearized Boltzmann equation 
in the collisionless limit using the random phase approximation. 
For this purpose, we add a small perturbation $U_{\sigma}({\bf q}, \omega)$ 
to evaluate the density response function. 
The linearized Boltzmann equation becomes 
\begin{eqnarray}
\frac{\partial f_{\sigma}^{0}}{\partial \varepsilon_{\sigma}^{0}}
\left \{
\left (
\omega - \frac{{\bf p}\cdot{\bf q}}{m} 
\right )
\nu_{\sigma}({\bf q}, {\bf p}, \omega)
+
\frac{{\bf p}\cdot{\bf q}}{m} 
\left [ 
U_{\sigma}({\bf q}, \omega)
+
g \delta n_{-\sigma}({\bf q}, \omega) 
\right ]
\right \}
= 0, 
\end{eqnarray}
where we neglect the collision integral on the right hand side. 
The fluctuation around static equilibrium is thus given by 
\begin{eqnarray}
\nu_{\sigma}({\bf r}, {\bf p}, \omega)
= 
-\frac{1}{\omega-\frac{{\bf p}\cdot{\bf q}}{m}}
\frac{{\bf p}\cdot{\bf q}}{m}
\left [
U_{\sigma}({\bf q}, {\bf p}, \omega)
+ 
g \delta n_{-\sigma}({\bf q}, \omega)
\right ]. 
\end{eqnarray}
Since the density fluctuation can be written as 
\begin{eqnarray}
\delta n_{\sigma}({\bf q}, \omega)
= 
\int \frac{d{\bf p}}{(2\pi\hbar)^{3}}
\frac{\partial f_{\sigma}^{0}}{\partial \varepsilon_{\sigma}^{0}}\nu_{\sigma}({\bf q}, {\bf p}, \omega), 
\label{RPA-1}
\end{eqnarray}
the density fluctuation in terms of a response function $\chi_{\sigma}^{0}({\bf q}, \omega)$ is 
given by 
$
\delta n_{\sigma}({\bf q}, \omega)
=
\chi_{\sigma}^{0}({\bf q}, \omega)
\left [
U_{\sigma}({\bf q}, \omega) + g \delta n_{-\sigma}({\bf q}, \omega)
\right ]
$, 
where 
the density response function $\chi_{\sigma}^{0}({\bf q}, \omega)$ is defined as 
\begin{eqnarray}
\chi_{\sigma}^{0}({\bf q}, \omega)
=
-\int \frac{d{\bf p}}{(2\pi\hbar)^{3}}
\frac{\partial f_{\sigma}^{0}}{\partial \varepsilon_{\sigma}^{0}}
\frac{1}{\omega-\frac{{\bf p}\cdot{\bf q}}{m}}
\frac{{\bf p}\cdot{\bf q}}{m}. 
\label{RPA-2}
\end{eqnarray}
We assume that the density perturbation is the same for the two components: 
$U({\bf q}, \omega) \equiv U_{\sigma}({\bf q}, \omega) = U_{-\sigma}({\bf q}, \omega)$.  
The density fluctuation is then reduced to 
$\delta n_{\sigma}({\bf q}, \omega) = \chi({\bf q}, \omega)U({\bf q}, \omega)$, 
where the response function is given by 
\begin{eqnarray}
\chi({\bf q}, \omega) 
= 
\frac{\chi_{\sigma}^{0}({\bf q}, \omega) [1+g \chi_{-\sigma}^{0}({\bf q}, \omega) ]}
{1-g^{2}\chi_{\sigma}^{0}({\bf q}, \omega)\chi_{-\sigma}^{0}({\bf q}, \omega)}. 
\end{eqnarray}

Zero of the denominator of a response function gives the frequency of the collective mode. 
In this case, the dispersion relation of the zero sound is obtained from the following equation:  
\begin{eqnarray}
1-g\chi^{0}({\bf q}, \omega)= 0, 
\label{EquationZeroSound}
\end{eqnarray} 
where we assumed the population balanced gas 
and used the relation $\chi^{0} ({\bf q}, \omega) \equiv \chi_{\sigma}^{0} ({\bf q}, \omega) 
= \chi_{-\sigma}^{0} ({\bf q}, \omega)$. 

Note that, this equation (\ref{EquationZeroSound}) at $T = 0$ reproduce the dispersion relation (\ref{ReZeroSound}). 
Solution of the linearized Boltzmann equation 
involves the denominator $\omega - {\bf p}\cdot{\bf q}/m$, 
as seen in Eq (\ref{RPA-2}). 
This means that excitations within the linearized Boltzmann equation can reproduce only the phonon regime: $\Omega \propto q$.

\end{document}